\documentclass[twocolumn,
 amsmath,amssymb,
 aps,
prb,
]{revtex4}

\usepackage{graphicx}
\usepackage{dcolumn}
\usepackage{bm}

\begin{document}

\title{
Electron transport in quantum wire superlattices
}

\author{Thomas Grange}
\email{thomas.grange@neel.cnrs.fr}
 \altaffiliation{Present address: Institut N\'{e}el--CNRS, 25 av. des Martyrs, 38042 Grenoble, France.}
\affiliation{
 Walter Schottky Institut, Technische Universit\"{a}t M\"{u}nchen, Am Coulombwall 3, D-85748, Garching, Germany
}

\date{\today}

\begin{abstract}
Electronic transport is theoretically investigated in laterally confined semiconductor superlattices using the formalism of non-equilibrium Green's functions.
Velocity-field characteristics are calculated
for nanowire superlattices of varying diameters, from the quantum dot superlattice regime
to the quantum well superlattice regime.
Scattering processes due to electron-phonon couplings, phonon anharmonicity, charged impurities, surface and interface roughness and alloy disorder are included on a microscopic basis.
Elastic scattering mechanisms are treated in a partial coherent way beyond the self-consistent Born approximation.
The nature of transport along the superlattice is shown to depend dramatically on the lateral dimensionality.
In the quantum wire regime, the
electron velocity-field
characteristics are predicted to deviate strongly from the standard Esaki-Tsu form. The standard peak of negative differential velocity is shifted to lower electric fields, while additional current peaks appear due to integer and fractional resonances with optical phonons.



\end{abstract}

\maketitle


\section{Introduction}

\begin{figure}
\begin{centering}
\includegraphics[width=0.45\textwidth]{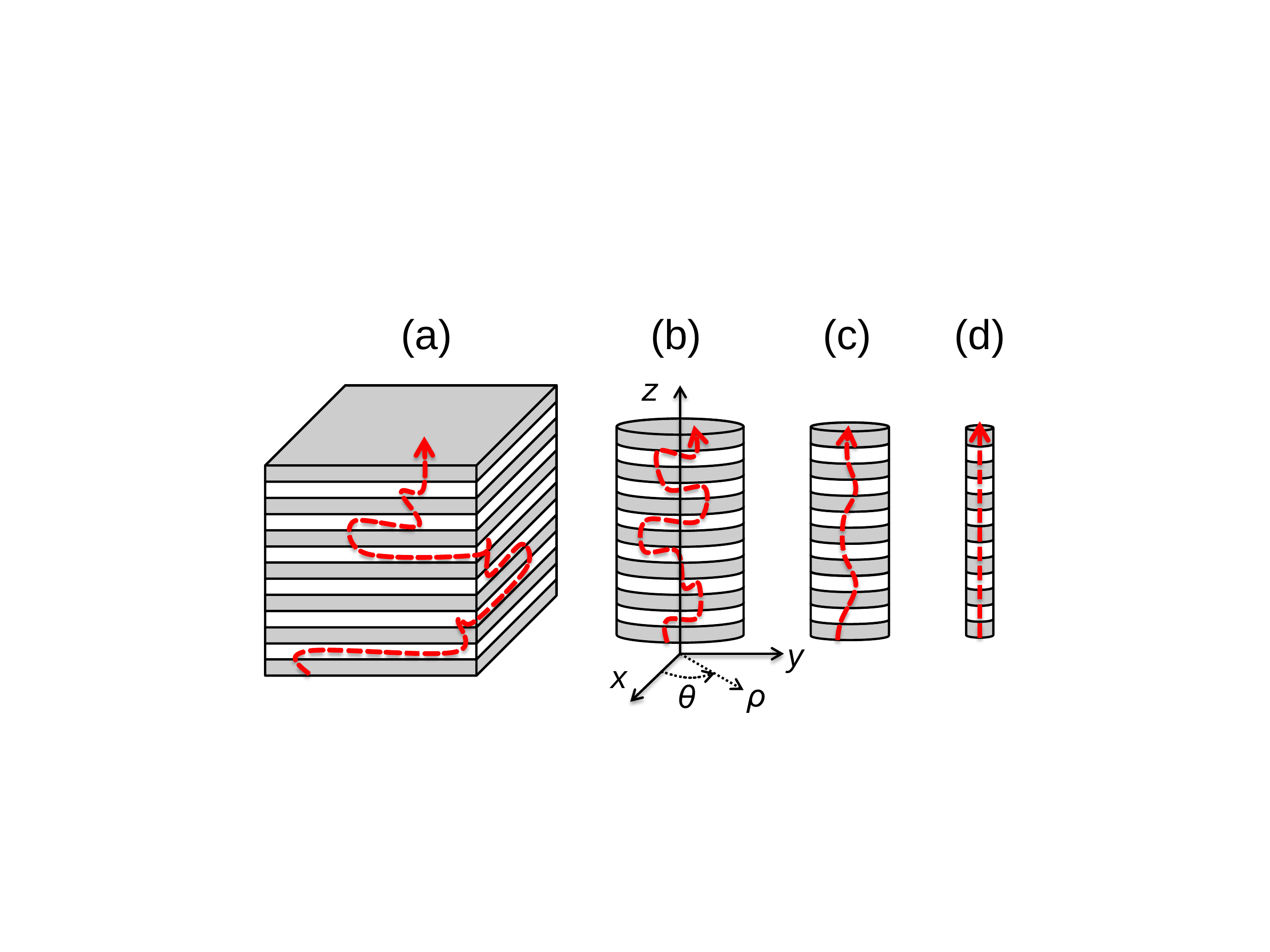} 
\end{centering}
\caption{(Color online). Schematic of superlattices with different lateral geometries. (a) shows a standard quantum well superlattice consisting of 2D ($x$,$y$) infinite layers of various materials stacked periodically along the $z$ direction. (b),(c) and (d) represent nanowire superlattices of different lateral sizes.
Dotted arrows are sketched of classical trajectories of charge carriers through the various superlattices under an electric field applied along the $z$ direction. Scattering processes occur due to various effects (interaction with phonons, structural disorder...). It illustrates the transition from transport involving 3D scattering processes towards a pure 1D motion with increasing quantum lateral confinement.
Cylindrical coordinates ($\rho$,$\theta$,$z$) used in the calculations are indicated for the (b) nanowire.
}
\label{SchematicSLs}
\end{figure}

Electron transport in superlattices (SLs) has been widely studied in the case of 1D periodic arrangement of 2D semiconductor layers 
\cite{esaki1970superlattice,sibille1990observation,wacker2002semiconductor} (Fig.~\ref{SchematicSLs}a).
In contrast, the nature of charge transport in superlattices of lower dimensionality
remains an open question.
The nature and efficiency of dissipative processes are well known to strongly depend on the dimensionality, with longest lifetimes and coherence times in 3D quantum confined structures \cite{bockelmann90, Benisty91, zibik09, grange2009decoherence}. Quantum dot (QD) SLs are hence expected to display radically different non-equilibrium transport properties than quantum well (QW) SLs.
Potential applications for high performance thermoelectric converters\cite{lin2003thermoelectric}, photovoltaics\cite{nozik2002quantum}, and quantum cascade lasers\cite{Wingreen97, grange2013nanowire} further motivates the fundamental understanding of the non-equilibrium transport properties of SLs with 3D quantum confinement.

Though ballistic transport in a QW SL is ideally a pure 1D problem, scattering processes couples all spatial degrees of freedom (Fig.~\ref{SchematicSLs}a).
In QW SLs, the in-plane free electron motion possesses a continuous dispersion. During the scattering processes, energy transfer occurs between the axial and lateral motions. 
Hence the effective energy conservation laws in the motion along the SL axis can greatly differ from the 3D one of the involved scattering mechanisms.
For example, in  QW heterostructures, 3D elastic scattering processes due to static defects
act effectively as 1D inelastic scattering mechanisms with respect to the motion along the SL.
In contrast, in the limit of purely 1D SL with 0D lateral motion (Fig.~\ref{SchematicSLs}d), the lateral state is frozen so that
the 3D energy conservation laws of the scattering processes apply directly to the 1D motion along the SL.

The effect of dissipation on 1D quantum electron transport has been investigated in finite quantum region such as atomic wires\cite{frederiksen2007inelastic} and stacked QDs \cite{gnodtke2006phonon}. Quantum dissipative transport in QD SLs has been studied in Ref.~\onlinecite{Vukmirovic07} for high electric fields. However full velocity-field characteristics of 1D SLs have not been calculated so far. In particular it is still unkown how the Esaki-Tsu negative differential velocity (NDV) \cite{esaki1970superlattice} evolves when the lateral dimensionality is reduced towards QD SLs.

In this article, we present a detailed theory of electron transport in semiconductor SLs laterally confined in nanowires using the framework of non-equilibrium Green's functions (NEGF). 
The formalism presented here allows to describe the transport in SLs for any lateral dimensionality from the regime of coupled 0D QDs (Fig.~\ref{SchematicSLs}d)
to the one of coupled 2D QWs (Fig.~\ref{SchematicSLs}a). 
For this purpose, we include in our modeling
the scattering mechanisms that are known to be important in
planar QW heterostructures
as well as the ones specific to quasi-0D systems.
Specifically, we consider the mechanism of anharmonic polaron decay which has been shown to be dominant in InGaAs self-assembled QDs \cite{grange2007polaron,zibik09}.
In addition, as the nature of elastic scattering mechanisms evolves from incoherent to coherent as the dimensionality decreases, we include
a coherent treatment of elastic scattering processes for lateral state-conserving processes. 
The electronic transport properties are shown to depend dramatically on the SL lateral dimensionality. With increasing lateral confinement, the drift velocity-voltage characteristics evolve from the Esaki-Tsu form for usual QW SL, dominated by a single NDV effect, towards richer characteristics where (i) the linear conductivity increases, (ii) the standard Esaki-Tsu NDV occurs at much lower voltages, and (iii) large peaks due to integer and fractional resonances with optical phonons appear.





The paper is organized as follows. The theory of electron transport in laterally confined SLs is described in section~\ref{theory}. The application to the transport in GaAs/AlGaAs nanowire superlattices of varying diameters is presented and discussed in section~\ref{results&discussions}.

\section{Theory}
\label{theory}
Our aim is to model the non-equilibrium electron transport in a nanowire SL heterostructure. 
We first introduce the general form of the Hamiltonian. 
We then introduce an electronic basis set, and present the NEGF formalism. We then treat the relevant scattering terms arising from electron-phonon interactions and static disorder effects.
Finally we present the method used for solving the self-consistent NEGF equations.



\subsection{Hamiltonian of the nanowire superlattice}

We consider charge carriers in a nanowire SL in interaction with phonon modes. 
The following Hamiltonian is used to model the system: 

\begin{equation}
\hat{H} = \hat{H}_0 + \hat{V}_{\text{e}} + \hat{V}_{\text{e-ph}} + \hat{H}_{\text{vib}} + \hat{V}_{\text{e-e}}^{\text{m-f}} ,
\end{equation}

where
$\hat{H}_0$ is the electronic Hamiltonian of a single charge carrier in the idealized heterostructure under a homogeneous electric field (the length gauge will be used throughout this paper);
$\hat{V}_{\text{e}}$ is the
static electronic potential due to disorder effects such as interface and surface roughness, charged impurities
and alloy disorder;
$\hat{V}_{\text{e-ph}}$ represents the interaction terms between electrons and phonons;
$\hat{H}_{vib}$ is the vibrational Hamiltonian of the crystal, including its anharmonic part;
$\hat{V}_{\text{e-e}}^{\text{m-f}}$ represents the mean-field potential arising from the Coulomb interaction with other charge carriers. 
The standard caret notation is used to refer to operators.

\subsection{Electronic structure and basis states}

We consider cylindrical nanowires of diameter $D=2R$ having SL heterostructures along their longitudinal axis.
We use cylindrical coordinates ($z,\rho, \theta$) where $z$ is the nanowire axis (see Fig.~\ref{SchematicSLs}b).
The material composition and modeling parameters are assumed to depend only on the $z$ coordinate.
The electronic structure is modeled considering a single band within the envelope function approximation. We consider a basis set of the form: 
\begin{equation}
\Psi_{\alpha,n}(z,\rho,\theta) = \zeta_{\alpha}(z) \phi_{n}(\rho,\theta) ,
\end{equation}
where the $\zeta_{i}(z)$ are defined below
and $\phi_{n}(\rho,\theta)$ are
eigenstates of the Schr\"{o}dinger equation on a homogeneous disk (i.e. the assumed nanowire cross-section):
\begin{equation}
\phi_{n}(\rho,\theta) = \frac{1}{\sqrt{\pi} R J_{({m_n}+1))}(\chi_{m_n}^{l_n})} J_{m_n}\left( \frac{\chi_{m_n}^{l_n} \rho}{R}\right) e^{i{m_n}\theta}
\end{equation}
where $J_m$ is the Bessel function of order $m$ and $\chi_m^l$ is its $l$th zero. The indexes $m_n$ and $l_n$ are chosen such that they are associated with the $n$th eigenvalue of the 2D lateral motion at the $z$ coordinate:
\begin{equation}
\mathcal{E}_{n} (z) =  \frac{\hbar^2 \left(\chi_{m_n}^{l_n}\right)^2}{2m^*(z) R^2}
\end{equation}
where $m^*(z)$ is the material-dependent effective mass.
For each lateral mode we are then left with a $z$-dependent electronic Hamiltonian:

\begin{equation}
\langle  n \vert \hat{H}_0 \vert p \rangle = \delta_{n,p} \left[ \hat{h}_0 - eF\hat{z} + \mathcal{E}_{n}(z) \right]  ,
\end{equation}

where 
$F$ is the electric field applied along the $z$ axis and
\begin{equation}
\hat{h}_0 = \frac{-\hbar^2}{2}\frac{\partial}{\partial z}\frac{1}{m^*(z)}\frac{\partial}{\partial z} + V_b(z),
\end{equation}
with $V_b(z)$ being the material-dependent band-offset potential.
The eigenstates of the periodic $\hat{h}_0$ Hamiltonian satisfy the Bloch theorem and can be classified by wavevector $\kappa_z \in [-\pi/l_p,\pi/l_p]$ and miniband index $\nu$
\begin{equation}
\hat{h}_0 \varphi^{\nu}_{\kappa_z}(z) = \varepsilon^{\nu}_{\kappa_z}\varphi^{\nu}_{\kappa_z}(z)
\end{equation}
In order to discretize the basis set in the $z$ direction, various choices are possible. A full real-space discretization\cite{lake97,kubis09} by e.g. finite differences requires a large basis size in order to quantitatively describe the miniband structure, rendering the NEGF numerical implementation challenging.
On the other hand, mode-space approaches are also possible,
and allow the description of the transport properties with a much smaller basis \cite{lee2002nonequilibrium,lee2006quantum,Vukmirovic07}.
Here, the Hilbert space is reduced by keeping only a finite number of minibands, i.e. the low lying states of the unbiased structure.
An energy cut-off $E_c$ is fixed and only the minibands separated from the ground state by less than $E_c$ are kept.
Formally we define the following projection operator
\begin{equation}
\hat{P}_z = \sum_{\nu,\kappa_z}^{\varepsilon^{\nu}_{0} - \varepsilon^{0}_{0} < E_{c}}  |\varphi^{\nu}_{\kappa_z}\rangle\langle \varphi^{\nu}_{\kappa_z} |
\end{equation}
In the following, the non-equilibrium dynamics will be computed within the subspace $\mathcal{P}$ generated by $\hat{P}_z$.
The various operators $\hat{O}$ defined on the full Hilbert space need to be transformed in operators $\hat{O}_r$ acting within the $\mathcal{P}$ subspace.
The restrictions of the $\hat{h}_0$ unperturbed Hamiltonian and $\hat{z}$ position operator are simply defined
by:
\begin{equation}
\hat{h}_{0r} = \hat{h}_0 \hat{P}_z
\end{equation}
\begin{equation}
\hat{z}_r = \hat{P}_z ~ \hat{z} ~ \hat{P}_z
\end{equation}
Within the $\mathcal{P}$ subspace,
we consider
the localized basis which is constructed by diagonalizing the $\hat{z}_r$ position operator \cite{lee2006quantum,Vukmirovic07}:
\begin{equation}
\hat{z}_r|\zeta_{\alpha}\rangle = \zeta_{\alpha} |\zeta_{\alpha}\rangle
\end{equation}
where the $|\zeta_{\alpha}\rangle$ ($\alpha=1,2,...$) are sorted
by increasing $\zeta_{\alpha}$ eigenvalues.
In addition, their phase is chosen in order to fullfill periodicity:
\begin{equation}
\zeta_{\alpha+pM}(z) = \zeta_{\alpha}(z-pL)
\end{equation}
where $p$ is an integer, $M$ is the number of minibands and $L$ is the period length.

All the terms $\hat{V}_{(j)}$ in the Hamiltonian other than $\hat{h}_0$ are functions of the sole $\hat{z}$ operator. In order to conserve the commutation relation $[\hat{V}_{(j)},\hat{z}] = [\hat{V}_{(j)r},\hat{z}_r] = 0$, \textit{i.e.} to conserve their local nature,
we use the following transformation:
\begin{equation}
\hat{V}_{(j)r} = \sum_{\alpha} |\zeta_{\alpha}\rangle\langle\zeta_{\alpha}| \hat{V}_{(j)} |\zeta_{\alpha}\rangle\langle\zeta_{\alpha}|
\label{Transform_Z_Dep_Hamilt_Terms}
\end{equation}
This transformation allows us to keep exactly the same form for the current operator \footnote{
In the work of Lee and coworkers \cite{lee2002nonequilibrium,lee2006quantum}, an implicit truncation of the Hamiltonian
leads instead to an additional current term $J_{\text{scatt}}$ which is found to be numerically negligible.
}:
\begin{equation}
\hat{j}_{zr} = \frac{i}{\hbar}[\hat{H}_r,\hat{z}_r] = \frac{i}{\hbar}[\hat{h}_{0r},\hat{z}_r]
\end{equation}



We also keep only a finite number of lateral states satisfying $\mathcal{E}_{i} - \mathcal{E}_{0} < E_c$.
In the following NEGF simulations, it will be checked that the $E_c$ cut-off value is chosen high enough so that adding higher minibands and/or higher lateral states does not change significantly the calculated current.
In the following, we will use exclusively operators acting in the $\mathcal{P}$ subspace and all the $r$ subscripts indicating restricted operators
will be omitted for simplicity.

\subsection{Non-equilibrium Green's functions}

\subsubsection{Electron Green's functions}

The NEGF formalism is generally introduced in terms of four Green's functions (GFs) \cite{datta1997electronic,haug2008quantum,caroli1971direct,mahan1987quantum}. Here the lesser, upper, retarded and advanced electron GFs are defined respectively in terms of the following quantities:

\begin{subequations}
\begin{equation}
G^<_{\alpha,\beta,n}(t_1,t_2) =   i \langle \hat{c}^+_{\beta,n}(t_2) \hat{c}_{\alpha,n}(t_1) \rangle
\label{lesser}
\end{equation}

\begin{equation}
G^>_{\alpha,\beta,n}(t_1,t_2)=   - i \langle \hat{c}_{\alpha,n}(t_1) \hat{c}^+_{\beta,n}(t_2) \rangle
\end{equation}

\begin{equation}
G^R_{\alpha,\beta,n}(t_1,t_2)=   -i \Theta(t_1-t_2) \langle \{ \hat{c}_{\alpha,n}(t_1) , \hat{c}^+_{\beta,n}(t_2) \} \rangle
\end{equation}

\begin{equation}
G^A_{\alpha,\beta,n}(t_1,t_2)=   i \Theta(t_2-t_1) \langle \{  \hat{c}_{\alpha,n}(t_1) , \hat{c}^+_{\beta,n}(t_2) \} \rangle
\end{equation}
\end{subequations}
where $\hat{c}^+_{\alpha,n}$ and $\hat{c}_{\alpha,n}$ are respectively the creation and annihilation operators in the electronic state $| \Psi_{\alpha,n} \rangle$.
It is convenient to introduce the GF operators $\hat{G}(t_1,t_2)$ defined here by:
\begin{equation}
\langle \alpha,n \vert \hat{G}(t_1,t_2) \vert \beta, n' \rangle = \delta_{n,n'}G_{\alpha,\beta,n}(t_1,t_2) , 
\end{equation}
where $G$ refers to any of the four GFs.
Since we do not consider spin-dependent effects, the spin index is not indicated here for simplicity and all quantities are implicitly spin diagonal.
The spectral function reads in operator form:
\begin{equation}
\hat{A} = i (\hat{G}^R - \hat{G}^A) =  i (\hat{G}^> - \hat{G}^<)
\label{spectral}
\end{equation}

The retarded GF can be expressed from the spectral function:
\begin{equation}
\hat{G}^R (t_1,t_2) = -i \Theta(t_1-t_2) \hat{A}(t_1,t_2)
\label{retarded-spectral}
\end{equation}

The lesser GF (\ref{lesser}) is an extension of the concept of density matrix for two different times. Its value at equal times corresponds to the density matrix: 
\begin{equation}
\hat{\rho}(t) = -i \hat{G}^<(t,t) .
\label{DensityMatrix}
\end{equation}
Here, off-diagonal terms of the GFs
are considered in the $z$ direction but not in the lateral directions where only the diagonal terms
of the $|\phi_{n}\rangle$ eigenmodes are considered. The calculations are thus independent of the basis choice along the $z$-axis, allowing full coherent description of transport along the nanowire.
In contrast, the use of only diagonal GF terms in the lateral directions
does not allow us to describe coherent transport effects in the lateral directions. This seems a reasonable approximation for thin nanowires where the lateral scattering potentials are weak compared to the lateral quantization energies. On the other hand, in the limit of large diameter nanowire and planar heterostructures, this is equivalent to considering only GF terms which are diagonal in the in-plane momentum, similarly to previous implementations of NEGF in planar quantum well heterostructures \cite{lake97,lee2002nonequilibrium,kubis09}. In this later case,
lateral localization effects might become relevant at low temperature and/or in presence of strong disorder effects, which remains beyond the scope of this work.

In steady state, the GFs depend only on the time difference $(t_2-t_1)$. In this case we will use one-time-dependent GFs $\hat{G}(t_1-t_2)=\hat{G}(t_1,t_2)$. Energy-dependent GFs are then defined by Fourier-transforming these quantities
\begin{equation}
\hat{G}(E) = \frac{1}{\hbar} \int dt \hat{G}(t) e^{iEt/\hbar}
\label{steady}
\end{equation}
Moreover, in steady state, the advanced and retarded GFs are linked by
\begin{equation}
\hat{G}^A(E)=\hat{G}^{R}(E)^+ ,
\end{equation}
so that the GFs are forming only two independent quantities.



\subsubsection{Phonon Green's functions}

The Hamiltonian of the vibration of the crystal reads

\begin{equation}
\hat{H}_{vib} = \hat{H}_{ph} + \hat{V}_{a}
\end{equation}

\begin{equation}
\hat{H}_{ph} = \sum \hbar \omega_{\mathbf{q},p} \hat{a}_{\mathbf{q},p}^+\hat{a}_{\mathbf{q},p}
\end{equation}
where $\hat{H}_{ph}$ and $\hat{V}_{a}$ are respectively the harmonic and anharmonic parts, and $\hat{a}_{\mathbf{q},p}^+$ and $\hat{a}_{\mathbf{q},p}$ are the creation and annihilation operators for a phonon mode of branch $p$ with wavevector $\mathbf{q}$.
The phonon GFs are defined by\cite{economou1984green}:

\begin{equation}
D^<_{\mathbf{q},p}(t_1,t_2) =   - i \langle \hat{A}^+_{\mathbf{q},p}(t_2) \hat{A}_{\mathbf{q},p}(t_1) \rangle
\end{equation}

\begin{equation}
D^>_{\mathbf{q},p}(t_1,t_2) =   - i \langle \hat{A}_{\mathbf{q},p}(t_1) \hat{A}^+_{\mathbf{q},p}(t_2) \rangle
\end{equation}

\begin{equation}
D^R_{\mathbf{q},p}(t_1,t_2)=   -i \Theta(t_1-t_2) \langle [  \hat{A}_{\mathbf{q}}(t_1) , \hat{A}^+_{\mathbf{q},p}(t_2) ] \rangle ,
\end{equation}

with
\begin{equation}
\hat{A}_{\mathbf{q},p} = \hat{a}_{\mathbf{q},p} + \hat{a}^+_{\mathbf{-q},p} .
\end{equation}
In steady-state, we will use the same one-time notation $D(t)$ and energy-dependent GFs $D(E)$ as for electronic GFs (Eq.~\ref{steady}). When only the harmonic part of the Hamiltonian is taken into account, the lesser and retarded equilibrium phonon GFs for a mode of frequency  $\omega$ read respectively:

\begin{equation}
D_{\omega}^{<(0)}(t) = - i [ (N_{\omega}+1) e^{i\omega t}  +
N_{\omega}
e^{-i\omega t} ]
\end{equation}



\begin{equation}
D_{\omega}^{R(0)}(t) = i \Theta(t) [ e^{i\omega t}  - e^{-i\omega t} ] ,
\end{equation}

where $N_{\omega}$ is the Bose factor at the energy $\hbar\omega$. In the energy domain, it reads:

\begin{equation}
D_{\omega}^{<(0)}(E) = -2\pi i [ (N_{\omega}+1) \delta(E+\hbar\omega)  +
N_{\omega}
 \delta(E-\hbar\omega)  ]
\end{equation}

\begin{equation}
D_{\omega}^{R(0)}(E) = \frac{1}{E+i0^+- \hbar\omega} - \frac{1}{E+i0^+ + \hbar\omega} .
\end{equation}

Harmonic phonon GFs are usually directly used in electron transport calculations. However, in this work, we will use instead anharmonic GFs for optical modes
which will be derived below (Sec. \ref{anharmonicity}).
This will allow us to account for the mechanism of anharmonic polaron decay.








\subsection{Equations of motion}


The equations of motion in the NEGF formalism can be expressed in terms of the so-called Dyson and Keldysh relations; in steady-state, they read respectively  \cite{datta1997electronic,haug2008quantum,caroli1971direct,mahan1987quantum}:
\begin{equation}
\hat{G}^R(E) = \left[ E\hat{I}-\hat{H}_0-\hat{V}_c - \hat{\Sigma}^R(E) \right]^{-1}
\label{Dyson}
\end{equation}
\begin{equation}
\hat{G}^<(E) = \hat{G}^R(E) \hat{\Sigma}^<(E) \hat{G}^A(E)
\label{Keldysh}
\end{equation}
where $\hat{V}_c$ is the part of the interaction which is treated coherently.
Here it reads $\hat{V}_c = \hat{V}_{\text {e-e}}^{\text{m-f}} + \hat{V}_e^{\text{rand}}$, where $\hat{V}_{\text {e-e}}^{\text{m-f}}$ is the mean-field Coulomb potential and $\hat{V}_e^{\text{rand}}$ is the part of the electronic disordered potential which is generated randomly and treated coherently (see below).
Similarly to the spectral GF, the spectral self-energy is defined as
\begin{equation}
\hat{\Gamma} = i (\hat{\Sigma}^R - \hat{\Sigma}^A) =  i (\hat{\Sigma}^< - \hat{\Sigma}^>) ,
\end{equation}

and the retarded self-energy can be expressed as:

\begin{equation}
\hat{\Sigma}^R(t) = -i\Theta(t) \hat{\Gamma}(t) = \Theta(t) (\hat{\Sigma}^<(t) - \hat{\Sigma}^>(t)) .
\end{equation}
In the energy domain, this relation reads:
\begin{equation}
\hat{\Sigma}^R(E) = -\frac{i}{2} \hat{\Gamma}(E) + \frac{1}{2}\mathcal{H}[\hat{\Gamma}](E)
\label{retarded-self-energy}
\end{equation}
where  $\mathcal{H}$ denotes the Hilbert tranform $ \mathcal{H}[\hat{\Gamma}](E) = \mathcal{P}\int \text{d}E' \hat{\Gamma}(E')/\pi(E-E')$.





\subsection{Self-energies due to electron-phonon interaction}

We consider the interactions of electrons with longitudinal optical (LO), surface optical (SO) and longitudinal acoustic (LA) phonon modes laterally confined in the nanowire.
Each interaction is of the form:
\begin{equation}
\hat{V}_{\text{e-ph}} = \sum_{\mathbf{q},p} \hat{f}_{\mathbf{q},p} \hat{A}_{\mathbf{q},p} .
\end{equation}
The form factors $\hat{f}_{\mathbf{q},p}$ are given in appendix \ref{app-phonons} for the different modes.
In the self-consistent Born approximation (SCBA), the self-energies due to electron-phonon interaction are of the form:
\begin{equation}
\hat{\Sigma}^{<}(t) = i \sum_{{\mathbf{q},p}}  \hat{f}_{\mathbf{q},p} \hat{G}^<(t) D_{\mathbf{q},p}^<(t)  \hat{f}_{\mathbf{q},p} .
\end{equation}

Within the 
$\{|\zeta_{\alpha}\rangle\}$ basis considered here, the nonvanishing coupling terms being diagonal with respect to the axial wavefunctions, the self-energy reads 
\begin{equation}
\Sigma^{<}_{\alpha \beta n}(t) = i \sum_{\mathbf{q},p} \sum_{n'} f^{(n,n')}_{\mathbf{q},p}(\alpha) f^{(n',n)}_{\mathbf{q},p}(\beta) G^<_{\alpha \beta n'}(t) D_{\mathbf{q},p}^<(t) ,
\end{equation}

where
\begin{equation}
f^{(n,n')}_{\mathbf{q},p}(\alpha) = \langle \Psi_{\alpha,n} \vert \hat{f}_{\mathbf{q},p} \vert \Psi_{\alpha,n'} \rangle.
\end{equation}


\subsubsection{Optical phonons}

For longitudinal optical (LO) and surface optical (SO) phonons, we assume wavevector-independent optical phonon GFs.
Indeed only long-wavelength optical phonons with negligible dispersion are effectively coupled to the relevant electronic states.
We can then write

\begin{equation}
\Sigma^{<\text{(iO)}}_{\alpha \beta n}(t) = i D_{\text{LO}}^<(t) \sum_{n'} W^{(n,n')}_{\text{iO}}(\alpha,\beta) G^<_{\alpha \beta n'}(t) ,
\label{SE-e-op-ph}
\end{equation}
where
\begin{equation}
W^{(n,n')}_{\text{iO}}(\alpha,\beta) = \sum_{\mathbf{q}} f^{(n,n')}_{\mathbf{q},\text{iO}}(\alpha) f^{(n',n)}_{\mathbf{q},\text{iO}}(\beta) .
\end{equation}
In the energy domain, it reads
\begin{equation}
\begin{split}
\Sigma^{<\text{(iO)}}_{\alpha \beta n}(E) & = i \sum_{n'} W^{(n,n')}_{\text{iO}}(\alpha,\beta) \\
& \times
\int \frac{dE'}{2\pi} D_{\text{LO}}^<(E')  G^<_{\alpha \beta n'}(E-E') .
\end{split}
\end{equation}
If the phonon GF $D_{\text{LO}}^<$ is taken harmonic, only quantized energy exchanges are allowed. 
In QDs, it has been shown that energy exchanges strongly differing from the optical phonon energy take place due to simultaneous electron-phonon interaction and anharmonic couplings among phonons \cite{grange2007polaron,zibik09}.
In these previous works, the electron-phonon interaction was diagonalized exactly, and the Fermi golden rule was used to calculate scattering rates among polaronic states induced by anharmonic couplings. Here, instead, the SCBA is used in order to treat the electron--optical-phonon interaction, and anharmonicity is included in the optical-phonon GFs. As already checked in Ref.~\onlinecite{Vukmirovic07}, the SCBA very well reproduces the exact polaron formation, especially for temperatures $T$ verifying $k_bT < E_{\text{LO}}$ where $E_{\text{LO}}$ is the optical phonon energy.
The calculation of anharmonic GFs is presented below.

\subsubsection{Acoustic phonons}
In contrast to optical phonons, the acoustic phonons are assumed to be harmonic and to have a linear dispersion.
The corresponding self-energy can be expressed as:
\begin{equation}
\Sigma^{<}_{\alpha \beta n}(t) = i
\sum_{n'}
K^{(n,n')}_{\alpha,\beta}(t)
G^<_{\alpha \beta n'}(t) ,
\end{equation}
where
\begin{equation}
K^{(n,n')}_{\alpha,\beta}(t) =  \sum_{\mathbf{q},\text{LA}}
D_{\omega_q}^{<(0)}(t)
f^{(n,n')}_{\mathbf{q},\text{LA}}(\alpha) f^{(n',n)}_{\mathbf{q},\text{LA}}(\beta) 
\end{equation}

\subsection{Green's functions of anharmonic phonons}
\label{anharmonicity}


%



We now calculate the anharmonic phonon GFs involved in the electron--optical-phonon self-energy.
We retain only the cubic couplings in the anharmonic terms, and thermal equilibrium population of phonon modes is assumed.
As we consider only diagonal terms in the phonon's GFs, the Dyson equation reads
\begin{equation}
D^R_{\mathbf{q},p}(E) = \frac{1}{\left( D_{\mathbf{q},p}^{R(0)}(E)\right)^{-1} - \Pi_{\mathbf{q},p}^R(E)} ,
\end{equation}

where $\Pi^R$ is the phonon retarded self-energy.
The lesser GF is then calculated using the Keldysh relation:
\begin{equation}
D_{\mathbf{q},p}^<(E) = D^R_{\mathbf{q},p}(E) \Pi_{\mathbf{q},p}^<(E)  D^A_{\mathbf{q},p}(E) .
\end{equation}
The phonon self-energy is calculated in the Born approximation, taking into account cubic anharmonic terms. The phonon lesser self-energy reads
\begin{equation}
\Pi^<_{\mathbf{q},p}(t)  = i \sum_{\mathbf{q_1},p_1,\mathbf{q_2},p_2} |V_{a} (\mathbf{q},\mathbf{q_1},\mathbf{q_2})|^2 D^{<(0)}_{\mathbf{q_1},p_1}(t) D^{<(0)}_{\mathbf{q_2},p_2}(t) ,
\end{equation}
in which we have taken the harmonic phonon GFs in the right hand side.
The cubic anharmonic coupling $V_{a}$ is expressed in terms of the Gr\"{u}neisen constant as reported in Ref.~\onlinecite{grange2007polaron}.
The phonon spectral and retarded self-energies are then given respectively by
\begin{equation}
\Gamma_{j}(E)
= i ( \Pi^<_{j}(-E) - \Pi^<_{j}(E) )
\end{equation}
\begin{equation}
\Pi^R_{j}(E)  = \frac{1}{2}\mathcal{H}(\Gamma_{j})(E) - \frac{i \Gamma_{j}(E)}{2}
\end{equation}
where  $\mathcal{H}$ denotes the Hilbert transform.
Applying the Dyson equation, we obtain the phonon lesser GF:


\begin{equation}
D_j^< (E) = \frac{4 E_j^2 \Pi^<(E)}{\left[ E^2-E_{j}^2 + E_j \mathcal{H}(\Gamma_{j})(E) \right]^2 + E_j^2 \Gamma_{j}^2 (E)} .
\end{equation}





\subsection{Elastic scattering}

Disordered potentials due to randomly distributed charged impurities, alloy disorder, rough interfaces and rough surfaces break the ideal symmetry of the structure, and couple the dynamics in the various directions. 
In particular, they couple the 1D motion along the $z$ superlattice axis to the the lateral motion.
In planar QW heterostructures, as the lateral dispersion forms a continuum, the 
static disorder
produces an incoherent and irreversible evolution within the $z$-electron basis.
In contrast, 
in the limit of purely 1D heterostructures, where only one lateral state is involved, the evolution is completely reversible, as elastic couplings induce a unitary evolution within the $z$-electron basis.
A treatment of elastic scattering beyond the SCBA is thus required in these low-dimensional structures.

In previous studies, the surface roughness in nanowires has been treated coherently by random generation of surfaces\cite{wang2005theoretical,lherbier2008quantum}. Here, we consider in addition interface roughness, impurity scattering and alloy scattering. Instead of generating microscopic random realizations of all surfaces, interfaces, charge positions, and alloy atom positions, we develop a simpler approach based on their correlations. We believe that it is sufficient for low or moderate disorder effects, which are anyway often not known precisely from the microscopic point of view.
The disordered potentials are split into diagonal and off-diagonal parts with respect to the lateral eigenstates:
\begin{equation}
\hat{V}_e^{\text{diag}} = \sum_{n} \langle \phi_n | \hat{V}_e | \phi_n \rangle ,
\end{equation}

\begin{equation}
\hat{V}_e^{\text{off-diag}} =  \sum_{n \neq m} \langle \phi_n | \hat{V}_e | \phi_m \rangle .
\end{equation}

The diagonal component is treated coherently, while the off-diagonal part can only be treated incoherently since we consider only diagonal elements of the GFs with respect to the lateral states.

\subsubsection{
Elastic scattering self-energies
}

The off-diagonal elastic coupling component $\hat{V}_e^{\text{off-diag}}$, which does not conserve the lateral state, is treated within the SCBA. The corresponding self-energy reads:




\begin{equation}
\Sigma^{<}_{\alpha \beta n}(t) = i \sum_{n'\neq n} \langle V_{nn'}(\alpha) V_{n'n}(\beta) \rangle G^<_{\alpha \beta n'}(t) ,
\end{equation}
where 
\begin{equation}
V_{nn'}(\alpha) = \langle \Psi_{\alpha,n} \vert \hat{V_e} \vert \Psi_{\alpha,n'} \rangle.
\end{equation}


\subsubsection{
Elastic scattering coherent terms}
\label{CoherentScat}

In order to treat coherently the diagonal component $\hat{V}_e^{\text{diag}}$, we first calculate the following covariance matrix
\begin{equation}
M_n(\alpha,\beta) = \langle V_{nn}(\alpha) V_{nn}(\beta) \rangle .
\end{equation}
We then generate random potentials with correlations that obey this calculated covariance. To this purpose, we need to rotate the covariance matrix in its principal component basis.
More precisely, the $M_n$ symmetric matrix can be diagonalized in
\begin{equation}
M_n = R_n D_n R^+_n ,
\end{equation}
where $D_n$ is a diagonal matrix and $R_n$ is a unitary matrix.
We then generate random diagonal matrices $d_n^{\text{rand}}$ in the following way: for each diagonal element of index $i$, we generate a $d_n(i,i)$ random number obeying a Gaussian distribution with a variance $D_n(i,i)$.
The random potential is then obtained by a back transformation into the original basis:

\begin{equation}
\hat{V}_e^{\text{rand}} = \sum_n | \phi_n \rangle \hat{V}_{e,n}^{\text{rand}}  \langle \phi_n | ,
\end{equation}
\begin{equation}
\hat{V}_{e,n}^{\text{rand}} = R_n d_n^{\text{rand}} R^+_n .
\end{equation}

The $\hat{V}^{\text{rand}}$ potential is then included in $V_c$ in the Dyson equation (\ref{Dyson}) in order to be treated coherently.
The all NEGF calculation is then made for different generated random potentials until the distribution of the calculated observables reached the desired accuracy.

\subsection{Field-periodic boundary conditions}

We assume the nanowire SL to contain an infinite number of periods of length $L$.
The possible formation of electrical field domains is not considered in this work\cite{esaki1974new,wacker2002semiconductor}.
Instead, the electric field is assumed to have the same periodicity as the SL.
The field-periodic boundary condition for the mean-field component of the electrostatic potential $V_{}^{\text{m-f}}$ reads
\begin{equation}
V^{\text{m-f}}(z+pFL) = V^{\text{m-f}}(z) ,
\label{Vperio}
\end{equation}
where $p$ is an arbitrary integer. Note that
the externally applied electric field $F$ is already included in $H_0$.
The Poisson equation reads
\begin{equation}
\frac{\partial^2 V_{}^{\text{m-f}}}{\partial z^2}  = -\frac{e}{\varepsilon_0 \epsilon_s} \left( 
\rho_e(z) + \rho_{d}(z)
\right),
\label{Poisson}
\end{equation}
where $\rho_e(z)$ and $ \rho_{d}(z)$ are the charge densities of carriers and dopants respectively.
From Eq.~\ref{Poisson}, this field-periodic boundary assumption implies that the electron probability distribution $\rho_e(z)$ is periodic.
In the length gauge used here, the periodicity condition for the GFs
reads
\begin{equation}
G_{\alpha+pM, \beta+pM,n}(E) = G_{\alpha, \beta, n}(E + pFL)
\label{Greenperio}
\end{equation}
where $M$ is the number of minibands.
Note that in nanowires, the SL periodicity can be broken by the inclusion of the coherent disordered potentials (\ref{CoherentScat}). In order to investigate such effects, the simulated period can be increased to several SL periods.
In this case, the period length $L$ in Eqs.~\ref{Vperio} and \ref{Greenperio} is replaced by $L_p = m_p L$ where $m_p$ is the number of SL periods included in one computational period. In order to know the needed number of periods, the integer $m_p$ is increased until the calculated observables tend towards constant values.


\subsection{Electron density and current}


The expectation values of the various observables can be derived using the relation between the density matrix and the lesser GF at equal times (Eq.~\ref{DensityMatrix}).
The electron probability distribution is given by:
\begin{equation}
\rho_e(z) = -2 i e   \sum_{\alpha,n}  \int \frac{\text{d}E}{2\pi} G^<_{\alpha \alpha n}(E) |\zeta_{\alpha}(z)|^2, 
\end{equation}
where the factor 2 arises from the spin index.
Using the expression of the current operator in terms of the restricted operators (Eq.~\ref{Transform_Z_Dep_Hamilt_Terms}),
the current reads
\begin{equation}
J_z = - \frac{2e}{\hbar V}    \sum_{\alpha,\beta}  \langle \beta\vert [\hat{h}_0,\hat{z}] \vert \alpha\rangle \sum_n \int \frac{\text{d}E}{2\pi}G^<_{\alpha \beta n}(E) .
\end{equation}


\subsection{Numerical details}


The self-consistent problem formed by the Keldysh equation, the Dyson relation, the expression of the various self-energies and the electrostatic Poisson equation is solved iteratively.
Starting from an initial guess of the GFs, we calculate iteratively
(i) the lesser $\hat{\Sigma}^<$ and upper $\hat{\Sigma}^>$ self-energies; (ii) the retarded self-energy (Eq.~\ref{retarded-self-energy}); (iii) the coherent part of the interaction, comprising the mean-field electrostatic potential and the random potential; (iv) the retarded GF from the Dyson equation; (v) the lesser GFs from the Keldysh relation.
This procedure is repeated until a self-consistent solution is reached, with two convergence criteria based on both the lesser GF and the calculated current.
Before each step (i), the GFs are renormalized in order to fulfill the charge neutrality for each period. It is checked that this renormalization factor converges accurately towards unity as the convergence is achieved.

The iterative steps (i) and (ii) are performed numerically in the time-domain. It is advantageous since the self-energies within the SCBA involves the product in the time domain. In contrast, energy domain calculations would involve numerically expensive convolutions, as a broad continuum of anharmonic optical phonons is considered.
On the contrary, the iterative steps (iv) and (v) are more easily computed in the energy domain, so that Fourier transforms of the GFs and the self-energies are used respectively in between steps (v)-(i) and (ii)-(iv).
Though adaptive energy grids have been shown to be useful in order to resolve GF peaks with a reduced number of energy points\cite{kubis09}, a homogeneous energy grid is used here in order to make use of a fast Fourier transform algorithm.

A maximum coherence length $l_c$ is set in the calculations, so that we consider only the GF terms $G_{\alpha, \beta, n}$ for $|\zeta_{\alpha}-\zeta_{\beta}| \leq l_c$.
Due to the field-periodic boundary conditions, the GFs $G_{\alpha, \beta, n}$ are calculated for $\alpha$ belonging to a given period, while $\beta$ varies away from $\alpha$ within the set coherence length $l_c$. In the simulation of the nanowire SL, we check that $l_c$ is taken large enough by increasing it until convergence of the calculated current.
In the regime of QW SL, we have verified that there is no difference in the calculated observables considering one or several periods, provided the doping densities remain low enough to prevent the formation of field domain instabilities.

\section{Results and discussions}
\label{results&discussions}
\subsection{Transport in nanowire SLs: transition from QW to QD SLs}

\begin{figure}
\begin{centering}
\includegraphics[width=0.45\textwidth]{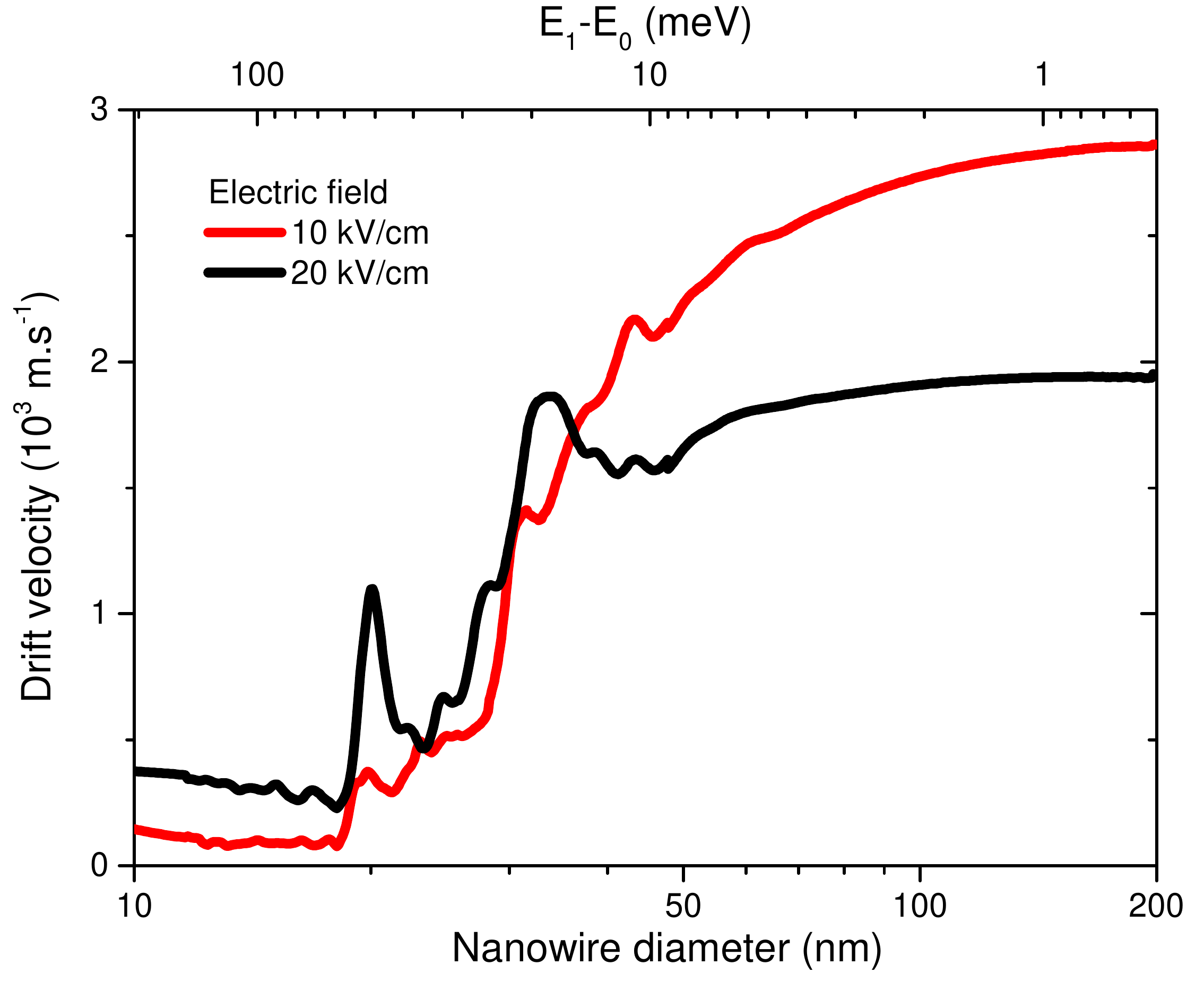} 
\end{centering}
\caption{(Color online). Drift velocity in a nanowire superlattice as a function of the nanowire diameter for various electric fields. The superlattice period consists of 5~nm GaAs followed by 5~nm AlGaAs. The temperature is set to 300~K.}
\label{IvsD}
\end{figure}

\begin{figure}
\begin{centering}
\includegraphics[width=0.45\textwidth]{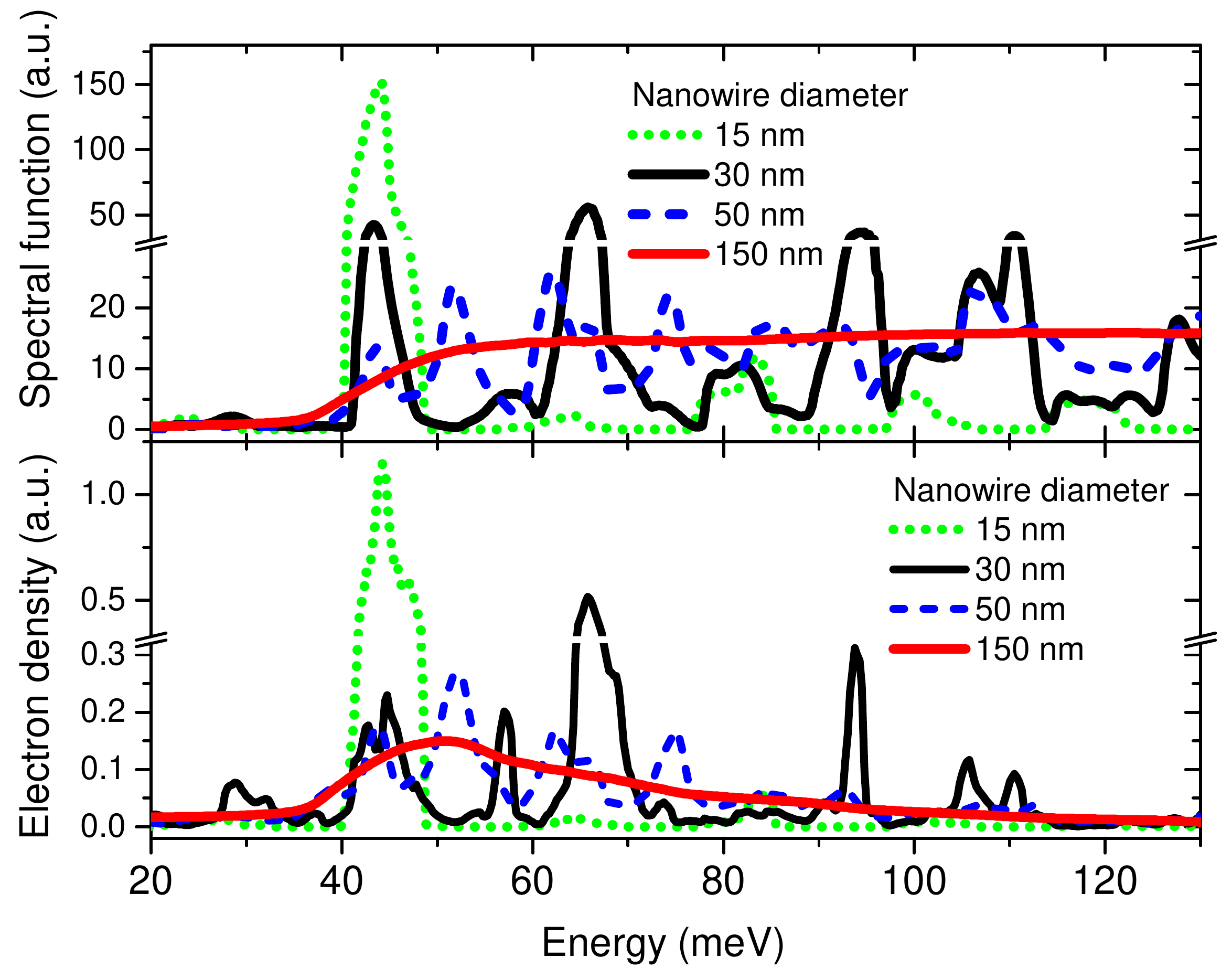} 
\end{centering}
\caption{(Color online). Spectral function $A_{\text{lat}}(E)$ (upper panel) and electron density $n_{\text{lat}}(E)$ (lower panel) for various nanowire diameters (see text for definitions). The electric field is 20~kV/cm and the temperature is 300~K.}
\label{Spectral}
\end{figure}

As an application of the theory presented above, we present calculations of the electronic transport in a GaAs/Al$_{0.3}$Ga$_{0.7}$As nanowire SL. One period consists in 5~nm GaAs (well) and 5~nm Al$_{0.3}$Ga$_{0.7}$As (barrier). The SL is assumed to be homogeneously n-doped with a concentration of $10^{16}$cm$^{-3}$. Assumed values for surface and interface roughnesses are given in appendix~\ref{app-elastic}.
The fundamental subband has a width of $9.7$~meV. It is separated from the first excited subband by a minigap of $122$~meV.
In the following calculations only the fundamental subband is considered (the cut-off value $E_c$ is taken around 100~meV).
Fig.~\ref{IvsD} shows the drift velocity as a function of the nanowire diameter for two different electric fields along the SL.
For large nanowire diameters, the current density tends toward a constant value. This is interpreted as the 2D regime for the lateral motion. In contrast, for smaller diameters, the quantum wire regime is reached, and the current density varies with the lateral confinement.
In order to confirm this interpretation, we plot in Fig.~\ref{Spectral} the quantities
\begin{equation}
A_{\text{lat}}(E) = \frac{1}{s} \sum_n A_{\alpha,\alpha,n}(E)
\end{equation}
\begin{equation}
n_{\text{lat}}(E) =-\frac{1}{s} \operatorname{Im}\left[ \sum_n G^<_{\alpha,\alpha,n}(E) \right],
\end{equation}
which represent respectively the spectral function and the electron density summed over the lateral states; $s=\pi R^2$ is the nanowire cross-section and $\alpha$ is the single basis state per period considered.
For a nanowire diameter of 150~nm, the smooth shape of the spectral function  indicates that the lateral quantization effects are overcome by broadening effects. Its flat shape is characteristic of a 2D subband.
On the other hand, for nanowire diameters below 17~nm, the current displays relatively small variations in Fig.~\ref{IvsD}. This corresponds to a quasi-0D regime for the lateral motion, where only the ground lateral state is occupied, as shown in Fig.~\ref{Spectral} for a nanowire diameter of 15~nm.
In between these two limit regimes, large variations of the current densities are calculated in Fig.~\ref{IvsD}, corresponding to an intermediate dimensional regime where several but still distinct lateral states are involved, as depicted in Fig.~\ref{Spectral} for nanowire diameters of 30 and 50~nm.

\begin{figure}
\begin{centering}
\includegraphics[width=0.45\textwidth]{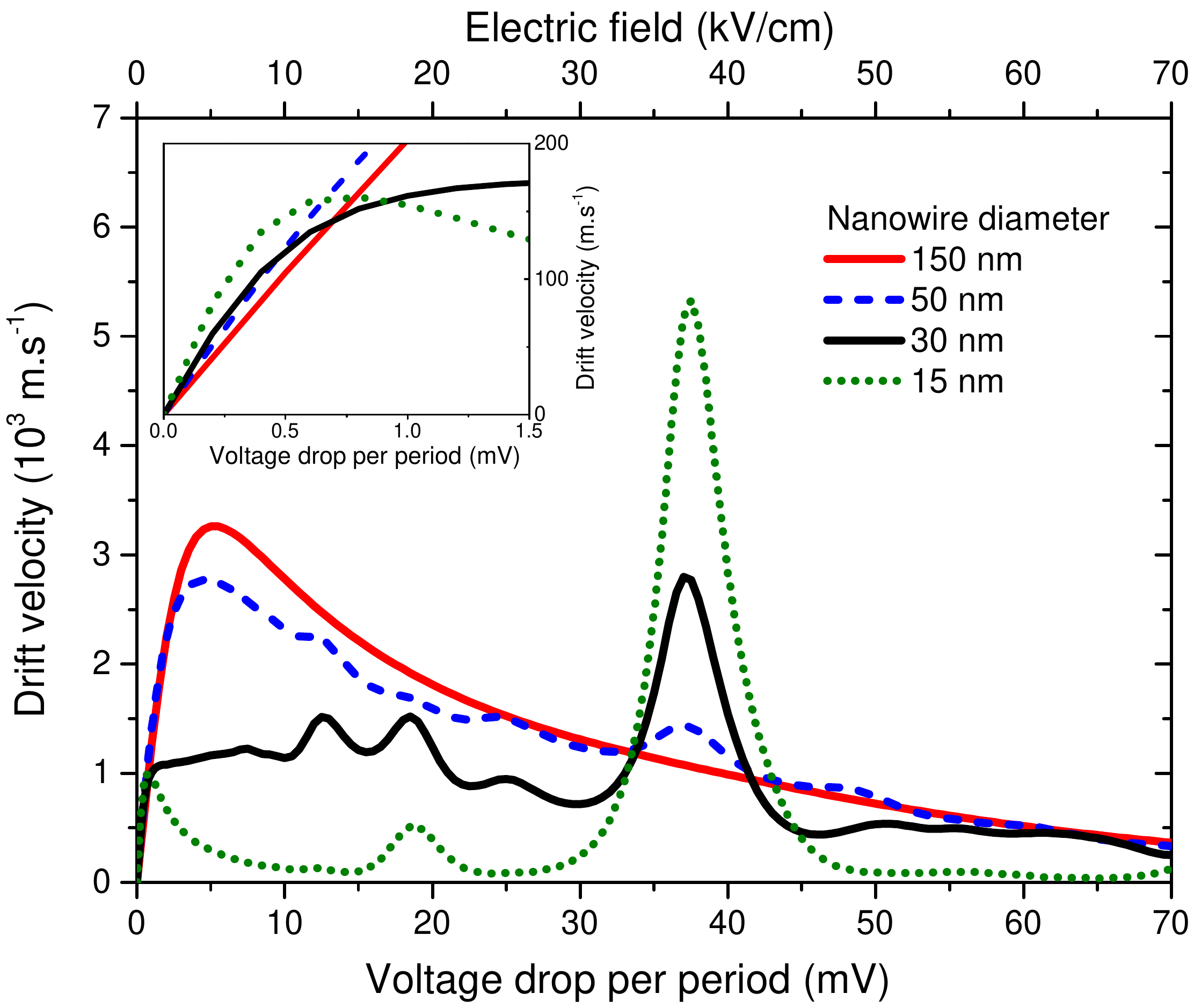}
\end{centering}
\caption{(Color online). Drift velocity--voltage characteristics at 300K for various diameters of the nanowire superlattice.}
\label{IVs}
\end{figure}

Fig.~\ref{IVs} shows the calculated velocity-voltage characteristics for various nanowire diameters.
For a nanowire diameter of 150~nm, where the 2D lateral regime is reached in good approximation, the standard Esaki-Tsu behavior is observed: negative differential velocity (NDV) is expected to occur when the carriers are excited beyond an inflection point in the miniband dispersion \cite{esaki1970superlattice}. 
For smaller diameters, the characteristics deviate from this simple standard behavior.
The most remarkable difference is the appearance of a large peak when the voltage drop per period matches the LO-phonon energy ($\hbar \omega_{\text{LO}}$=37 meV in GaAs). In addition, smaller resonances become visible at $\hbar\omega_{\text{LO}}/2$ and $\hbar\omega_{\text{LO}}/3$.
The interpretation is the following: in the 2D regime, the lateral motion acts as a continuous energy reservoir, which renders the 3D energy conservation laws of the scattering mechanisms not directly visible on the SL transport properties. In particular, though LO phonons provide the most efficient inelastic scattering processes, their quasi-monochromatic nature is not evidenced on the velocity--voltage characteristics. In contrast, in the limit of 1D SLs with 0D lateral motion, the energy is exchanged directly between the quantized SL levels (i.e. the Wannier-Stark states) and the phonons.
Hence resonances in the transport occur when two consecutive Wannier-Stark levels are separated by the LO-phonon energy $\hbar \omega_{\text{LO}}$. The smaller resonances around $\hbar\omega_{\text{LO}}/2$ and $\hbar\omega_{\text{LO}}/3$ are attributed to LO-phonon assisted transport between states distant of 2 and 3 periods respectively.

In the inset of Fig.~\ref{IVs}, we observe an increase in the electron mobility with decreasing NW thickness. This is attributed to the decrease of the efficiency of the scattering processes with decreasing dimensionality.
In addition, the position of the first NDV peak shifts to smaller voltage values with decreasing NW thickness.
This can be understood within the Esaki and Tsu semi-classical model~\cite{esaki1970superlattice}, where the NDV peak is predicted to occur for $edF\tau/\hbar=1$ ($\tau$ being the scattering time). Beyond this critical point, the Esaki-Tsu model predicts that electrons are accelerated beyond an inflection point in the energy--momentum dispersion relation and hence experience a negative effective mass.
This $edF=\hbar/\tau$ critical point can also be interpreted as the transition between the domain of validity of miniband transport and Wannier-Stark hopping models \cite{wacker2002semiconductor}.
In both pictures, the overall reduction of scattering processes with decreasing dimensionality explains qualitatively the observed shift of the NDV peak towards lower voltages.

\subsection{Role and nature of elastic scattering processes}

\begin{figure}
\begin{centering}
\includegraphics[width=0.45\textwidth]{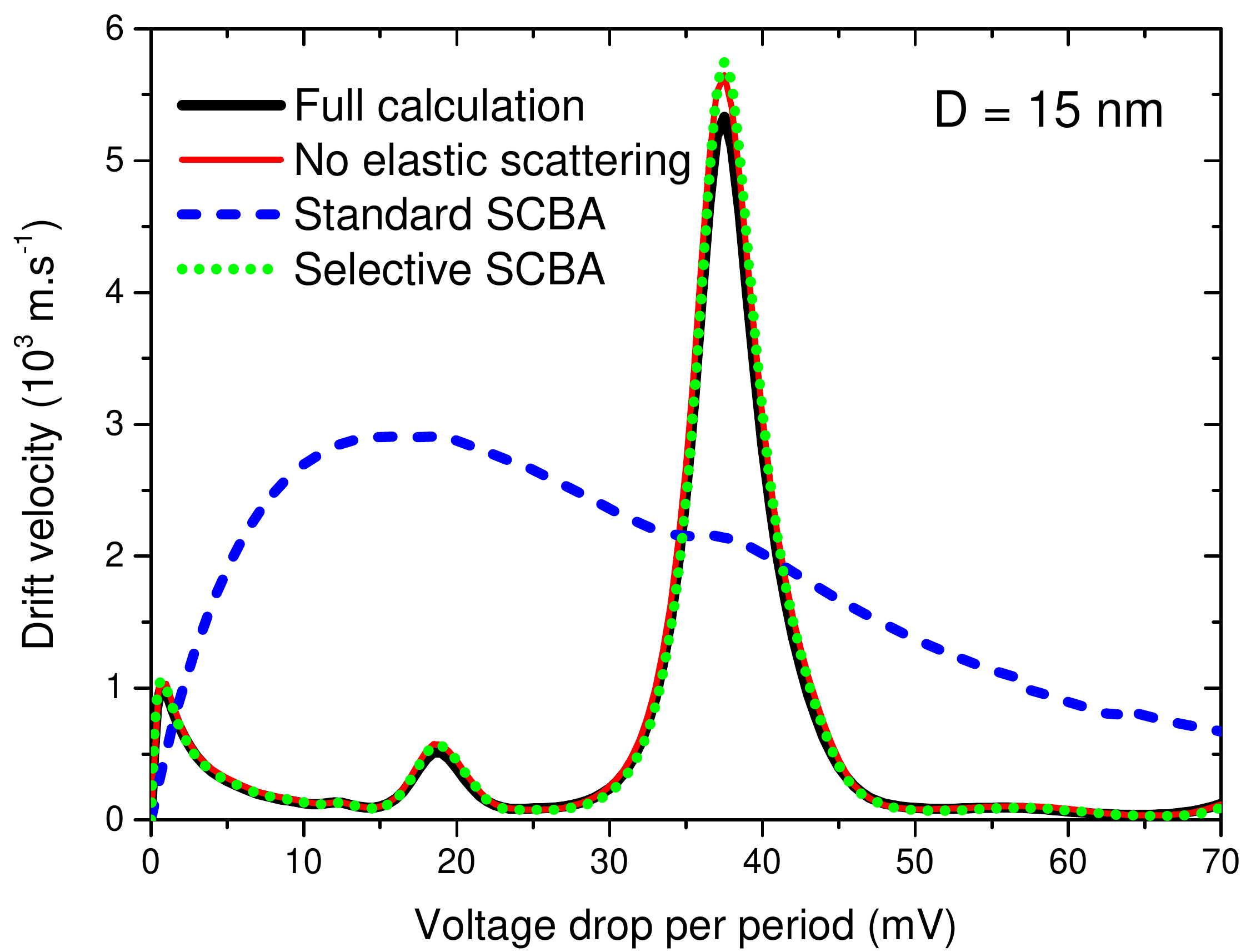}
\includegraphics[width=0.45\textwidth]{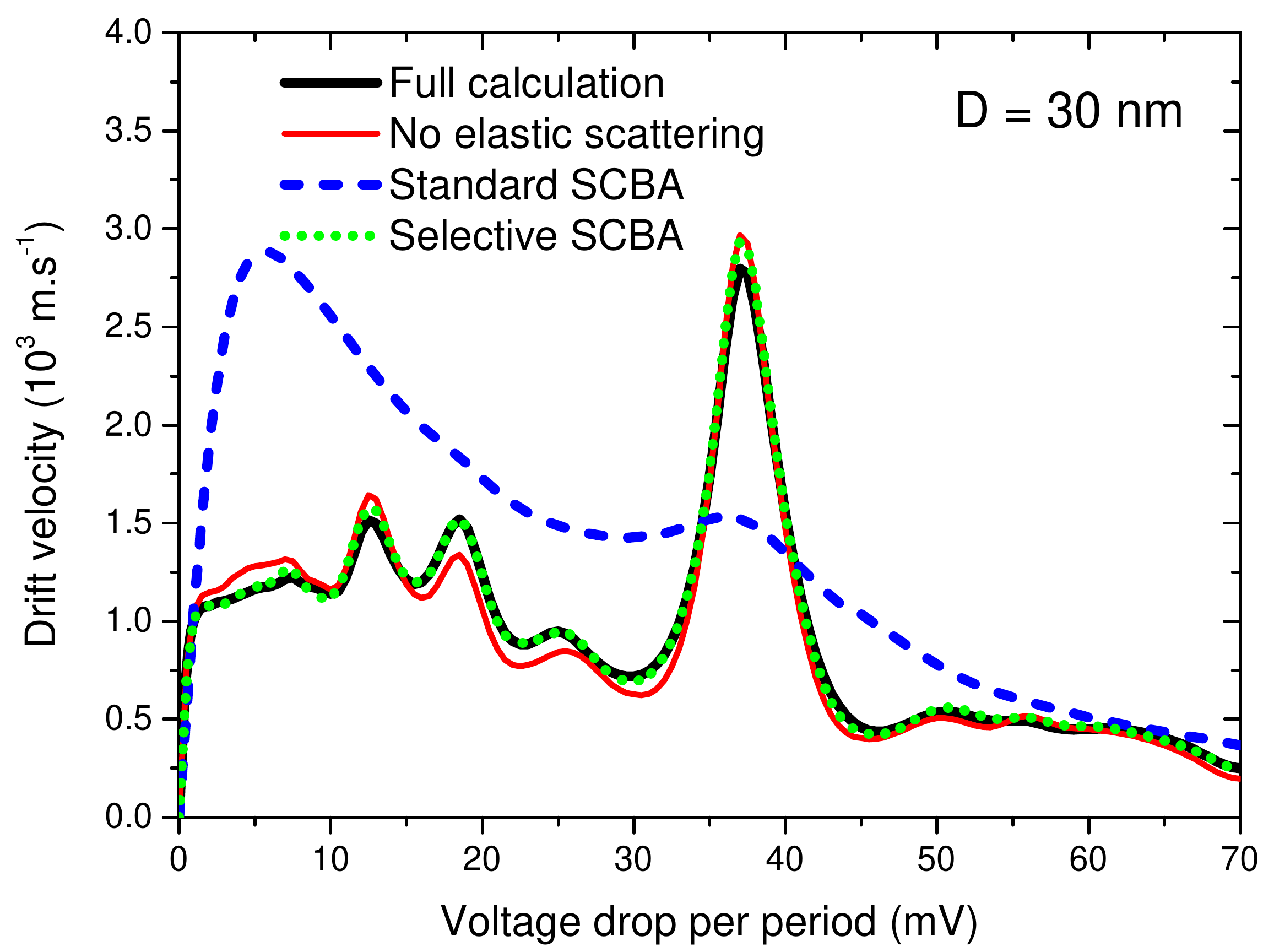}
\includegraphics[width=0.45\textwidth]{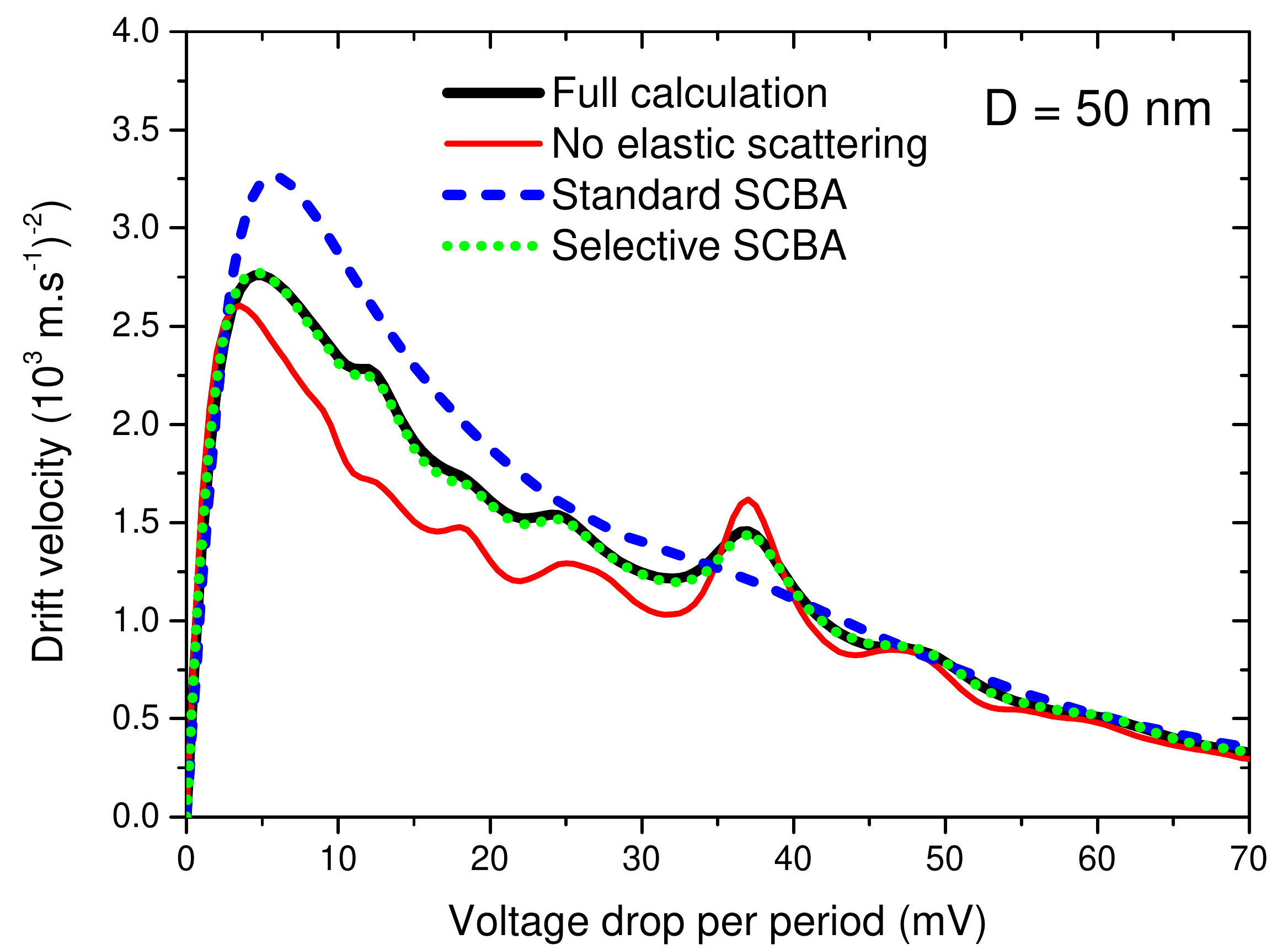}
\includegraphics[width=0.45\textwidth]{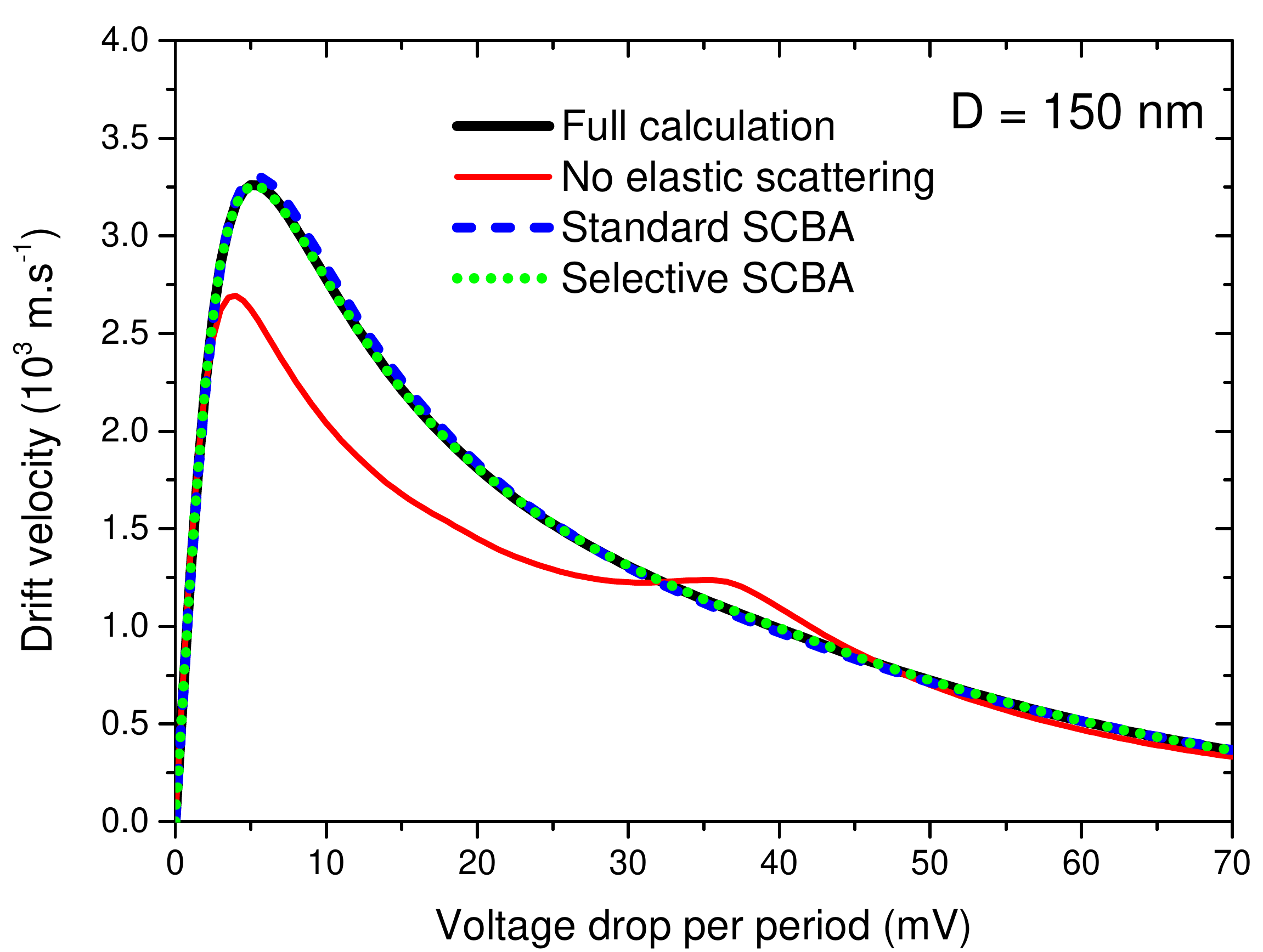}
\end{centering}
\caption{(Color online). Drift velocity--voltage characteristics at 300K for different treatments of the scattering mechanisms.}
\label{ComparisonScattering}
\end{figure}

In order to analyze the role and the nature of elastic scattering processes, simulations with different treatments of elastic scattering are shown in Fig.~\ref{ComparisonScattering}.
First, we discuss the comparison between full calculations with the one neglecting elastic scattering. 
These two calculations notably differ for thick NWs. In contrast, for thin NWs, the relative difference in the calculated transport is very small, indicating that elastic scattering only plays a minor role.
This strong reduction of elastic scattering processes with decreasing dimensionality is consistent with the intuitive expectation that elastic mechanisms are suppressed by the discretization of energy levels.

Now we compare the calculations using the standard SCBA, the ``selective SCBA'' defined above and the full calculation.
In the QW limit the standard SCBA, the ``selective SCBA" and the full calculation are found to give the same results. In contrast, in the quantum wire regime, the standard SCBA totally differs from the two other models.
The reason is that the standard SCBA fails
when the fluctuations of the energy levels become larger than their linewidths.
In contrast, in the present full calculation, the main coupling elements, which are diagonal with respect to the lateral eigenstates, are treated coherently.
If we now compare the full calculation to the selective SCBA calculation, we can see that the selective SCBA provides an excellent approximation to the full calculation, down to 15~nm thin nanowires.
Our interpretation is that the typical energy fluctuations of the levels stay smaller than the miniband width and are too weak to induce localization effects.




\subsection{Role of phonon anharmonicity}

\begin{figure}
\begin{centering}
\includegraphics[width=0.45\textwidth]{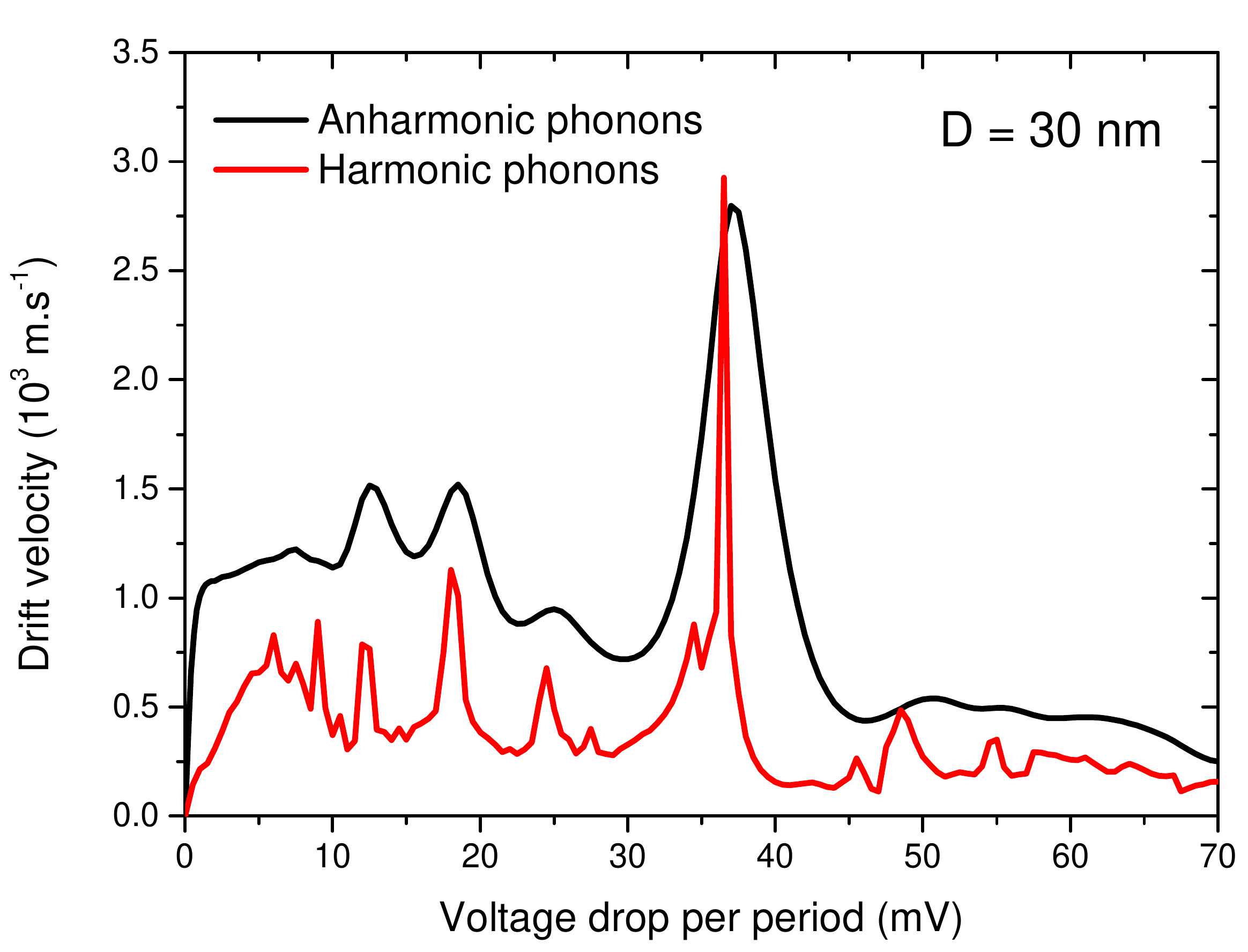}
\includegraphics[width=0.45\textwidth]{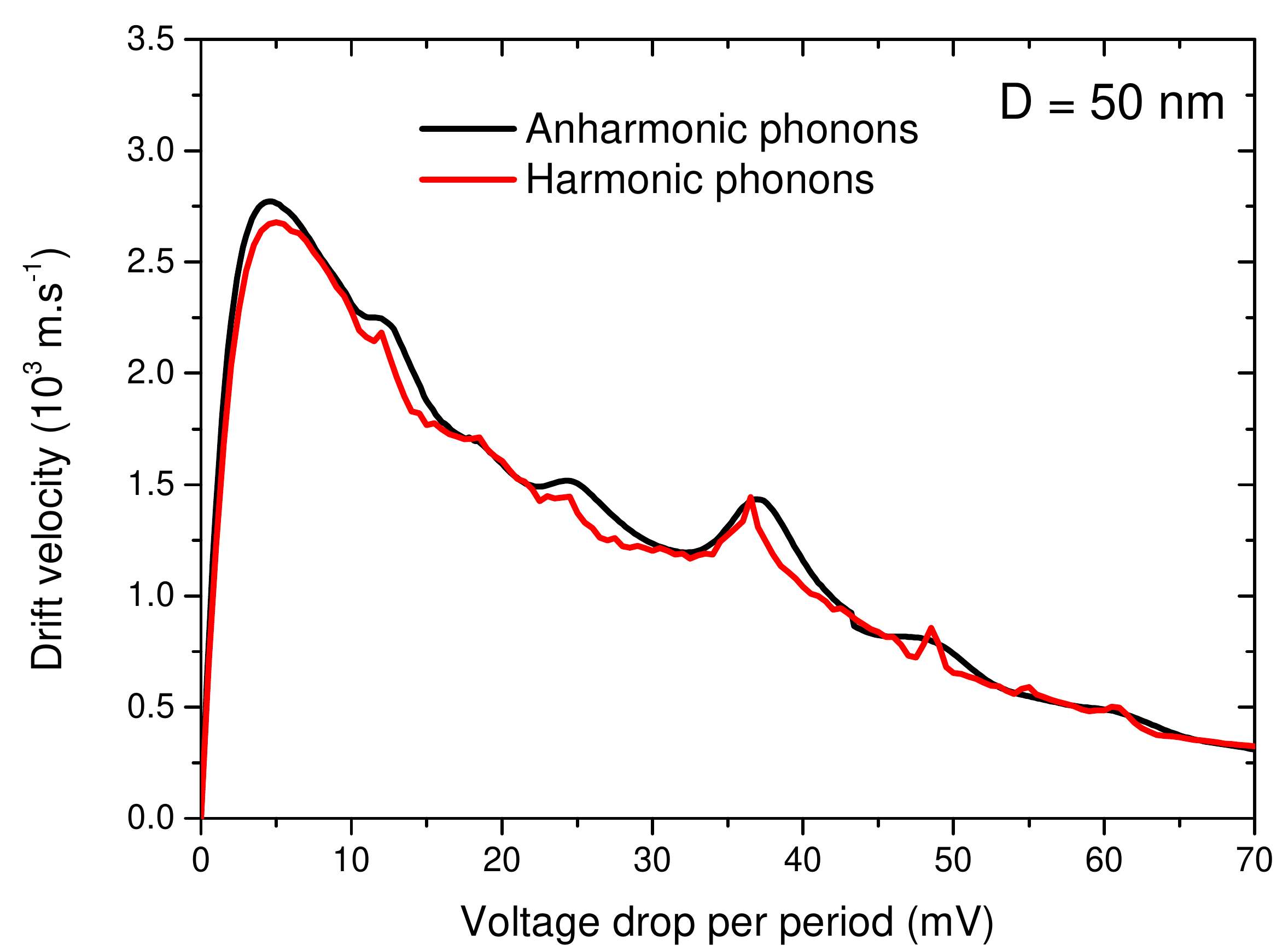}
\includegraphics[width=0.45\textwidth]{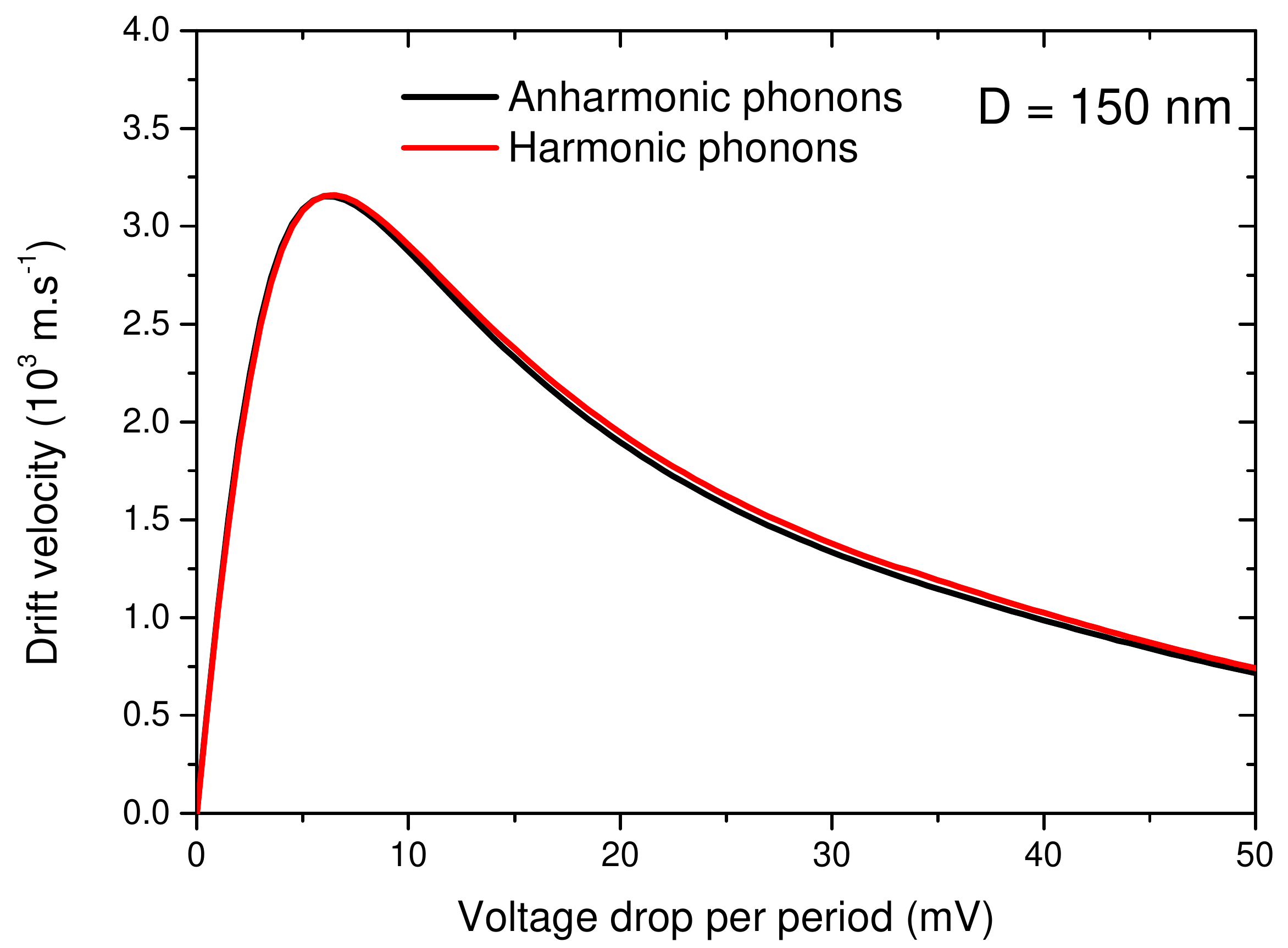}
\end{centering}
\caption{(Color online). Drift velocity--voltage characteristics at 300K with and without the inclusion of phonon anharmonicity.}
\label{Harmonicity}
\end{figure}

In Fig.~\ref{Harmonicity} we show a comparison of the calculations with and without the inclusion of phonon anharmonicity for various NW diameters.
While the inclusion of anharmonicity is found to be almost irrelevant for thick NWs in the QW SL limit, striking differences are observed for thin NWs.
Indeed the direct LO-phonon emission is gradually suppressed with decreasing dimensionality towards 0D systems, while the mechanism of anharmonic polaron decay becomes dominant.
For very thin NW diameters corresponding to the pure 1D SL limit, note that the simulation without phonon anharmonicity does not converge due to the strong reduction of the linewidths below our numerical possibilities in energy resolution.
Hence our calculations show a clear transition in the dominant energy dissipation pathway as the lateral dimensionality is reduced, from direct LO-phonon emission to polaron anharmonicity.

\section{Conclusion}
Within the NEGF formalism, we have developed a theory of electronic transport in nanowire SLs.
This model allows the calculation of transport in a wide range of NW SL thickness, for lateral dimensionality ranging from 0D to 2D, i.e. from QD SL to QW SL.
We have studied how the transport evolves with this change of lateral dimensionality. While velocity-voltage characteristics are dominated by the standard Esaki-Tsu NDV in QW SLs, electron-phonon resonances appear when the lateral quantum regime is reached. In addition, the electron mobility increases and the NDV peaks are strongly shifted to lower electric fields.
This is accompanied by important changes in the dominant scattering mechanisms.
In QW SLs, the transport is mainly controlled by LO-phonon scattering and elastic scattering processes. In contrast, when the lateral dimensionality is reduced, these two mechanisms are progressively suppressed, and anharmonic polaron decay becomes dominant.
\section{Acknowledgments}
This work has been supported by the Alexander von Humboldt foundation and the Austrian Science Fund FWF through Project No. F25-P14 (SFB IR-ON). S. Birner, M. Branstetter, C. Deutsch, P. Greck, G. Koblm\"{u}ller, M. Krall, T. Kubis, S. Rotter, K. Unterrainer, and P. Vogl are gratefully acknowledged for fruitful discussions.


\appendix

\section{Electron-phonon interactions in cylindrical nanowires}

\label{app-phonons}

In this appendix we give the form factor of the couplings of electrons with longitudinal-optical (LO), surface-optical (SO), and longitudinal-acoustical (LA) phonon modes confined in a cylindrical nanowire.
We consider the full confinement of phonons by the nanowire surfaces but neglect the effect of the SL interfaces.

\subsection{Interaction with longitudinal optical modes}

The polar interaction between electrons and optical phonons confined in a cylindrical nanowire is treated within the dielectric continuum model\cite{xie2000bound}.
The zone-center LO phonons involved in the couplings with electrons are assumed to be monochromatic with frequency $\omega_{\text{LO}}$.
 The interaction with the longitudinal optical (LO) modes of the nanowire reads


\begin{equation}
\hat{H}_{\text{e-LO}} = \sum_{m,l,q_z} f^{\text{LO}}_{m,l,q_z} (\rho,\theta,z)	\hat{A}^{\text{LO}}_{m,l,q_z} ,
\end{equation}

where

\begin{equation}
\hat{A}^{\text{LO}}_{m,l,q_z}  = \hat{a}^+_{m,l,q_z} + \hat{a}_{-m,l,-q_z}
\end{equation}
is the sum of phonon creation and annihilation operators with opposite momentum.
The form factor reads

\begin{equation}
f^{\text{LO}}_{m,l,q_z}(\rho,\theta,z)  = C^{\text{LO}}_{m,l,q_z} J_m\left( \frac{ \chi^l_m \rho }{R} \right) e^{im\phi} e^{iq_z z} ,
\end{equation}

where
\begin{equation}
|C^{\text{LO}}_{m,l,q_z}|^2 = \frac{e^2\hbar \omega_{\text{LO}}}																														{2\pi \varepsilon_0 L J_{m+1}^2(\chi_m^l) (\chi_m^{l 2} + R^2q_z^2)} \left( \frac{1}{\varepsilon_{\infty}} - \frac{1}{\varepsilon_{s}} \right) .
\end{equation}
$\varepsilon_0$ is the vacuum permittivity; $\varepsilon_{s}$ and $\varepsilon_{\infty}$ are respectively the static and high frequency relative permittivities.

\subsection{Interaction with surface optical modes}
The interaction with the surface optical modes (SO) reads\cite{xie2000bound}

\begin{equation}
\hat{H}_{\text{e-SO}} = \sum_{m,l,q_z} C^{\text{SO}}_{m,l,q_z} I_m(q_z\rho) e^{im\theta} e^{iq_z z} \hat{A}^{\text{SO}}_{m,l,q_z} ,
\end{equation}

where


\begin{equation}
\begin{split}
 |C^{\text{SO}}_{m,l,q_z}|^2 & = \frac{e^2\hbar \omega_{m,l,q_z}}																														{2\pi \varepsilon_0 L  I_m(q_z R)  (I_{m-1}(q_z R) + I_{m+1}(q_z R))} \\ &
\times  \frac{1}{q_z R}  \left( \frac{1}{\epsilon_{\infty}-\epsilon}-\frac{1}{\epsilon_{s}-\epsilon}  \right)
\end{split}
\end{equation}

and

\begin{equation}
\epsilon = - \frac{I_m(q_z R) ( (K_{m-1}(q_z R) + K_{m+1}(q_z R) ) }
{K_m(q_zR) ( (I_{m-1}(q_z R) + I_{m+1}(q_z R) ) }
\epsilon_{\text{ext}}
\end{equation}

where $\epsilon_{\text{ext}}$ is the relative permittivity outside the nanowire.

\begin{equation}
\omega_{m,l,q_z}^2 = \left( 1 + \frac{\epsilon_0 - \epsilon_{\infty}}{\epsilon_{\infty}-\epsilon} \right) \omega^2_{\text{TO}}
\end{equation}









\subsection{Interaction with acoustic phonons}
We consider the deformation potential interaction with acoustic phonons confined laterally in the nanowire.
In principle, the boundary condition couple longitudinal acoustic (LA) and transverse acoustic (TA) modes.
Coupled LA-TA modes interacting with electrons
have been considered by Yu \textit{et al} \cite{yu1995electron}. However, there, only phonon modes with axial symmetry have been considered. We expect this approximation to be valid in the thin nanowire limit but to completely fail for large nanowires where LA phonons do not necessarily propagate parallel to the nanowire axis.
Here, instead, we neglect the coupling between LA and TA modes (i.e. between compressive and shear modes) at the surface of the nanowire, and consider the interaction of electrons with all the LA modes.
This treatment is exact in the bulk limit and is expected to provide a reasonable approximation for nanowires of moderate size, as far as the NW diameter is large compared to the lattice constant. We believe such assumption for the boundary conditions does not change significantly the results for the electron transport properties, as (i) the nanowire thicknesses studied here are relatively large compared to the atomic scale and (ii) the efficiency of the acoustic phonon scattering mechanism is always much weaker than other scattering mechanisms.
The displacement fields for purely compressive phonon modes can be written as
\begin{equation}
u = \nabla \phi
\end{equation}
We consider the free boundary condition, so that $\nabla u =0$ at the surface and $\phi$ can be written as
\begin{equation}
\phi = \sum_{m,l,q_z} C_{m,l,q_z} J_m\left( \frac{ \chi_m^l \rho }{R} \right) e^{im\phi} e^{iq_z z}
\label{phi_phonon}
\end{equation}

The phonon quantization reads
\begin{equation}
\hat{u} =  \sum_{m,l,q_z} u_{m,l,q_z} \hat{A}^{\text{LA}}_{m,l,q_z} ,
\end{equation}
with the normalization condition:
\begin{equation}
\int d^3 r |u_{m,l,q_z}(r)|^2 = \frac{\hbar}{2\mu \omega_{m,l,q_z}} ,
\end{equation}
$\mu$ is the material density and $\omega_{m,l,q_z}$ the phonon frequency.

The integration using the form of Eq.~\ref{phi_phonon} gives:
\begin{equation}
|C_{m,l,q_z}|^2  = \frac{\hbar}{2 \pi L \mu \omega (\chi_m^{l 2} + R^2q_z^2) J_{m+1}^2(\chi_m^l)}
\end{equation}

The Hamiltonian describing the electron--LA-phonon interaction is taken of the form


\begin{equation}
\hat{H}_{\text{e-LA}} = D ~ \nabla.\hat{u}  ,
\end{equation}
where $D$ is the deformation potential of the band of interest. Here it gives:

\begin{equation}
\hat{H}_{\text{e-LA}} = D (\chi_m^{l 2}/R^2 + q_z^2) \hat{\phi} ,
\end{equation}


and finally:
\begin{equation}
\hat{H}_{\text{e-LA}} = \sum_{m,l,q_z} f^{\text{LA}}_{m,l,q_z} (\rho,\theta,z)	\hat{A}^{\text{LA}}_{m,l,q_z} ,
\end{equation}

where the form factor reads:

\begin{equation}
f^{\text{LA}}_{m,l,q_z}(\rho,\theta,z)  = C^{\text{LA}}_{m,l,q_z} J_m\left( \frac{ \chi_m^l \rho }{R} \right) e^{im\phi} e^{iq_z z} ,
\end{equation}
\begin{equation}
|C^{\text{LA}}_{m,l,q_z}|^2 = \frac{D^2 \hbar \sqrt{ \chi _m^{l 2} + R^2q_z^2}}																														{2 \pi R  L \mu c_s J_{m+1}^2(\chi_m^l)} ,
\end{equation}
where  $c_s$ the sound celerity.
The dispersion relation reads 
\begin{equation}
\omega_{m,l,q_z} = c_s \sqrt{ \frac{\chi _m^{l 2}}{R^2} + q_z^2 } .
\end{equation}



\section{Elastic scattering mechanisms}
\label{app-elastic}

Imperfections and intrinsic atomic disorder break the structure symmetry and the crystal periodicity, thus responsible for elastic scattering processes.
We consider below the effects of surface roughness of the nanowires, interface roughness of the superlattice heterostructures, randomly located charged impurities and alloy disorder.


\subsection{Interface roughness}

We note $z_{i}^0$ the mean $z$-coordinate of the $i$th interface. The deviation from this average position at the in-plane coordinate $\mathbf{r}$ is denoted by $\delta z_i(\mathbf{r})$.
The interface roughness can be statistically characterized by (i) the distribution of the interface position along the heterostructure growth axis, and (ii) its in-plane autocorrelation. All interfaces are assumed to obey the same statistics.
We note $P(z)$ the distribution of the $\delta z_i$ interface deviation 
and $\sigma$ its root mean square. We note $C(r)$ the in-plane autocorrelation form factor so that:

\begin{equation}
\langle \delta z_i(\mathbf{r}_1) \delta z_i(\mathbf{r}_2) \rangle = \sigma^2 C(|\mathbf{r}_2-\mathbf{r}_1|)
\end{equation}
We note $\Delta V$ the heterostructure band offset. The coupling correlations involved in the calculation of elastic scattering processes read:

\begin{equation}
\langle V^{\text{(ir)}}_{n n'}(\alpha) V^{\text{(ir)}}_{n' n}(\beta) \rangle =
\Delta V^2
Y_{n n'}
\sum_i W^{(i)}_{\alpha \beta} ,
\end{equation}
where
the lateral form factor reads
\begin{equation}
Y_{n n'} = \int d\mathbf{r}_1 d\mathbf{r}_2 \phi_{n}(\mathbf{r}_1) \phi_{n'}(\mathbf{r}_2) C(|\mathbf{r}_2-\mathbf{r}_1|) ,
\end{equation}
and the axial form factor for the $i$th interface reads
\begin{equation}
W^{(i)}_{\alpha \beta} = \int_{z^0_{i-1}}^{z^0_{i+1}} \text{d}z_i P(z_i-z_i^0) w^{(i)}_{\alpha}(z_i) w^{(i)}_{\beta}(z_i) ,
\end{equation}
with
\begin{equation}
w^{(i)}_{\alpha}(z_i) = \int\limits_{[z_{i}^0,z_{i}]} \text{d}z \vert \zeta_{\alpha}(z)\vert^2 .
\end{equation}

In the numerical study, $P(z)$ is taken as a Gaussian distribution with a standard deviation $\sigma = 0.15$~nm and the autocorrelation $C(r)$ decreases as an exponential $e^{-r/\lambda}$ with a correlation length $\lambda=8$~nm.

\subsection{Surface roughness}


We consider small fluctuations of the surface around its ideal cylindrical position.
The surface of the nanowires are considered as infinite barriers for the electrons. Motivated by the fact that a homogeneous change $\delta R$ in the nanowire diameter would induce a change $ \frac{\partial \mathcal{E}_{n}}{ \partial R} \delta R$ in the $n$-th lateral state energy, we make the following assumption for the surface roughness couplings \cite{sakaki1987interface}:

\begin{equation}
\langle n | V^{\text{(sr)}}(\theta,z) | n' \rangle = 
\sqrt{\frac{\partial \mathcal{E}_{n}}{\partial R}
\frac{\partial \mathcal{E}_{n'}}{\partial R}}
\delta R_{\theta,z}
\end{equation}


We assume an exponential autocorrelation of the form
\begin{equation}
\begin{split}
\langle \delta R(\theta_1,z_1) \delta R(\theta_2,z_2) \rangle & = \delta R^2 e^{-|z_2-z_1|/L_c^s} \\
& \times e^{-R|\arg [e^{i(\theta_2-\theta_1)}]|/L_c^s}
\end{split}
\end{equation}
in which we have factorized the axial and azimuthal correlation terms in order to simplify the calculations.
Finally we find that the correlations of the coupling terms are given by
\begin{equation}
\langle V^{\text{(sr)}}_{n n'}(\alpha) V^{\text{(sr)}}_{n' n}(\beta) \rangle =
\frac{4\delta R^2}{R^2}
\mathcal{E}_{n}
\mathcal{E}_{n'}
Y_{nn'}^{\text{(sr)}} W_{\alpha \beta}^{\text{(sr)}} ,
\end{equation}
where the lateral form factor reads

\begin{equation}
\begin{split}
Y_{n n'}^{\text{(sr)}} & = 
\int_{-\pi}^{\pi} \frac{\text{d}\theta}{2\pi} e^{i(m_n-m_{n'})\theta} e^{-R|\theta| /L_c^s}\\
& =
\begin{cases}
\frac{L_c^s}{\pi R} \left[ 1 - e^{-\pi R /L_c^s} \right] & \text{if } m_n=m_{n'} \\
\frac{R \left[ 1 - e^{- \pi R /L_c^s}(-1)^{m_n-m_{n'}} \right] }{\pi L_c^s \left[(m_n-m_{n'} )^2+(R/L_c^s)^2\right]} 
& \text{if } m_n \neq m_{n'}  \\
\end{cases} ,
\end{split}
\end{equation}

and the axial form factor reads
\begin{equation}
W^{\text{(sr)}}_{\alpha \beta} = \int \text{d}z \int  \text{d}z' \vert \zeta_{\alpha}(z)\vert^2 \vert \zeta_{\beta}(z')\vert^2 e^{-|z-z'|/L_c^s} .
\end{equation}

In the numerical study, the standard deviation is taken as $\delta R = 0.15$~nm and the surface correlation length is taken as $L_c^s=8$~nm.







\subsection{Charged impurities}
The random location of charged impurities is also a source of elastic scattering.
For simplicity of the calculation and for clear comparison with the 2D limit, we assume that the filling factor of the nanowires is unity.
The Coulomb potential created by ionized impurities at the position $\mathbf{r_d}$ reads
\begin{equation}
V^{\text{(c)}}_{\mathbf{r_d}}(\mathbf{r}) = \frac{e^2 e^{-(|\mathbf{r}-\mathbf{r_d}|)/L_s}}{4\pi \varepsilon_0 \varepsilon_s |\mathbf{r}-\mathbf{r_d}|}
\end{equation}
where $L_s$ is the screening length.
Following Nelander and Wacker \cite{nelander2009temperature}, $L_s$ is calculated within the Debye screening model for a 3D homogeneous electron gas with the same average electron density and at the lattice temperature. 
After performing an in-plane Fourier transform this potential reads \cite{haug2009quantum}: 

\begin{equation}
V^{\text{(c)}}_{\mathbf{\rho_d},z_d}(\mathbf{\rho},z) = \int \frac{d^2\mathbf{k_{\parallel}}}{4\pi^2} e^{i \mathbf{k_{\parallel}}.\rho} \frac{e^2}{2\epsilon_0 \varepsilon_s \epsilon_s k_{s}} e^{-k_s|z-z_d|}
\end{equation}
with $k_{s}=\sqrt{k_{\parallel}+1/L_s^2}$.
The coupling correlation terms are obtained by averaging over the different possible position of the charged impurities:
\begin{equation}
\begin{split}
\langle V^{\text{(c)}}_{n n'}(\alpha) V^{\text{(c)}}_{n' n}(\beta) \rangle_{z_d,\rho_d} =
\int dz_d \int d^2\rho_d ~ N_d(z_d) \\
\times
\langle \Psi_{\alpha,n} | V^{\text{(c)}}_{z_d,\rho_d} | \Psi_{\alpha,n'} \rangle
\langle \Psi_{\beta,n'} | V^{\text{(c)}}_{z_d,\rho_d} | \Psi_{\beta,n} \rangle
\end{split} .
\end{equation}

After some algebra we find: 
\begin{equation}
\langle V^{\text{(c)}}_{n n'}(\alpha) V^{\text{(c)}}_{n' n}(\beta) \rangle = 
\frac{\pi e^4}{2  \varepsilon_0^2 \varepsilon_s^2}
\int_{0}^{+\infty} \frac{k_{\parallel}dk_{\parallel}}{k_s^2}  F^2_{k_{\parallel}}(n,n') \Lambda_{k_{\parallel}}(\alpha,\beta) ,
\end{equation}

where $F_{k}$ are Hankel transform coefficients:


\begin{equation}
F_{k}(n,n') = \int_0^{R}  d\rho ~ \rho ~ \phi_n(\rho) \phi_{n'}(\rho)  J_{m_{n,n'}}\left( k \rho \right) ,
\end{equation}
with $m_{n,n'} =|m_n-m_n'|$, and the axial form factor reads:
\begin{equation}
\Lambda_{k}(\alpha,\beta) =
\int dz_1 \int dz_2 |\zeta_{\alpha}(z_1)|^2  |\zeta_{\beta}(z_2)|^2 \lambda_{k}(z_1,z_2), 
\end{equation}
with
\begin{equation}
\lambda_{k}(z_1,z_2)  =  \int dz N_d(z) e^{-k(|z_2-z|+|z_1-z|)} .
\end{equation}

\subsection{Alloy disorder}

We consider a ternary alloy of the form $A_xB_{1-x}C$. We assume a random and uncorrelated distribution of the atoms $A$ and $B$.
We assume that, at the atomic scale, the carriers experience local band offset potentials $V_{AC}$ or $V_{BC}$ with probability $x$ and $1-x$, respectively \cite{bastard1990wave, lake97}. Here these potentials are taken as the one of the binary compounds.
The mean potential reads
\begin{equation}
\overline{V}(z) = x V_{AC} + (1-x) V_{BC} ,
\end{equation}
while the alloy scattering potential is defined by:
\begin{equation}
\Delta V = V_{AC}-V_{BC}
\end{equation}



The variance of the local band offset
is then:
\begin{equation}
\begin{split}
\langle (V-\overline{V})^2 \rangle & = x(V_{AC} - \overline{V})^2 + (1-x)(V_{BC} - \overline{V})^2 \\
& = x(1-x) \Delta V^2
\end{split}
\end{equation}

We denote $\mathcal{V}$ the volume occupied by a binary pair of atoms ($\mathcal{V} = a^3/4$ for zinc-blende structures, $a$ being the lattice constant).
The alloy disorder coupling terms read
\begin{equation}
\langle V^{\text{(a)}}_{n n'}(\alpha) V^{\text{(c)}}_{n' n}(\beta) \rangle =
\mathcal{V} \Delta V ^2 A_{\alpha \beta} R_{n n'}
\end{equation}
where $x(z)$ is  the $z$-dependent alloy composition, and where the axial and lateral form factors read respectively:

\begin{equation}
A_{\alpha \beta} = \int_{\text{Alloy}} dz  |\zeta_{\alpha}(z)|^2 |\zeta_{\beta}(z)|^2 x(z)[1-x(z)] ,
\end{equation}

\begin{equation}
R_{n n'} = \int d^2\rho |\phi_{n}(\rho)|^2 |\phi_{n'}(\rho)|^2 .
\end{equation}

\end{document}